\documentclass[aps,superscriptaddress,showpacs,twocolumn,dvips]{revtex4}
\usepackage{feynmp}
\usepackage{amssymb}
\usepackage{epsfig}
\usepackage{graphicx}
\usepackage{subfigure}
\usepackage{mathrsfs}
\usepackage{hyperref}
\usepackage{cases}
\usepackage{bbm}
\usepackage{CJK}
\usepackage{xcolor}

\usepackage{lmodern}
\usepackage{textcomp}

\begin{document}


\title{Role of four-fermion interaction and impurity in the states of two-dimensional semi-Dirac materials}

\date{\today}

\author{Jing Wang}
\altaffiliation{E-mail address: jing$\textunderscore$wang@tju.edu.cn}
\affiliation{Department of Physics, Tianjin University, Tianjin 300072, P.R. China}
\affiliation{Department of Modern Physics, University of Science and
Technology of China, Hefei, Anhui 230026, P.R. China}

\begin{abstract}
We study the effects of four-fermion interaction and impurity on the low-energy states
of two-dimensional semi-Dirac materials by virtue of the unbiased renormalization group
approach. The coupled flow equations that govern the energy-dependent evolutions of all
correlated interaction parameters are derived after taking into account
one-loop corrections from the interplay between four-fermion interaction and impurity.
Whether and how four-fermion interaction and impurity influence the low-energy properties
of two-dimensional semi-Dirac materials are discreetly explored and addressed attentively.
After carrying out the standard renormalization group analysis, we find that both trivial
insulating and nontrivial semimetal states are qualitatively stable against
all four kinds of four-fermion interactions. However, while switching on both four-fermion
interaction and impurity, certain insulator-semimetal phase transition and the distance of
Dirac nodal points can be respectively induced and modified due to their strong interplay
and intimate competition. Moreover, several non-Fermi liquid behaviors that deviate from
the conventional Fermi liquids are exhibited at the lowest-energy limit.


\end{abstract}


\maketitle


\section{Introduction}

Dirac fermions have been becoming one of most significant subjects in contemporary condensed matter
physics~\cite{Novoselov2005Nature,Neto2009RMP,Kane2007PRL,Roy2009PRB,Moore2010Nature,Hasan2010RMP,
Qi2011RMP,Sheng2012Book,Bernevig2013Book} in the last decade and attract a vast of both experimental~\cite{Novoselov2005Nature,Neto2009RMP,Hasan2010RMP,Qi2011RMP,
Sheng2012Book,Bernevig2013Book} and theoretical efforts~\cite{Neto2009RMP,Kane2007PRL,Roy2009PRB,
Moore2010Nature,Hasan2010RMP,Qi2011RMP,Sheng2012Book,Bernevig2013Book}. These Dirac systems harbor
a variety of types, such as two-dimensional (2D) graphene~\cite{Neto2009RMP}, Weyl~\cite{Neto2009RMP,Burkov2011PRL,Yang2011PRB,Savrasov2011PRB,
Huang2015PRX,Weng2015PRX,Hasan2015Science,Hasan2015NPhys,Ding2015NPhys},
Dirac~\cite{WangFang2012PRB,Young2012PRL,Steinberg2014PRL,Hussain2014NMat,
LiuChen2014Science,Ong2015Science} semimetals and etc., which usually possess
several discrete Dirac nodal points with gapless low-energy excitations
and display a linear dispersion in two or three directions~\cite{Neto2009RMP,Hasan2010RMP,Qi2011RMP,
Huang2015PRX,Hasan2015Science,Ding2015NPhys}. Recently, some groups elucidate~\cite{Hasegawa2006PRB,Pardo2009PRL,Katayama2006JPSJ,Dietl2008PRL,Delplace2010PRB,Wu2014Expre}
that there exist another kind of analogous material, dubbed as the 2D semi-Dirac (SD) material,
which is of remarkable interest 2D Dirac-like material as its dispersion is parabolic in one
direction and linear in the other, such as the $\mathrm{VO_2-TiO_2}$ multilayer systems
(nanoheterostructures)~\cite{Pardo2009PRL}, quasi-two dimensional organic conductor
$\alpha-(\mathrm{BEDT-TTF)_2I_3}$ salt under uniaxial pressure~\cite{Katayama2006JPSJ},
tight-binding honeycomb lattices for the presence of a magnetic field~\cite{Dietl2008PRL},
and photonic systems consisting of a square array of elliptical dielectric cylinders~\cite{Wu2014Expre}.
These unique and unusual properties of the low-energy excitations~\cite{Banerjee2009PRL,Delplace2010PRB,Saha2016PRB}
may result in distinct behaviors from the general Dirac fermions in the low-energy regime~\cite{Neto2009RMP,Hasan2010RMP,Qi2011RMP,Huang2015PRX, Hasan2015Science,Ding2015NPhys}.
Recently, K. Saha~\cite{Saha2016PRB} study this 2D SD system and indeed found that
some unconventional properties can be engineered. For instance, an electromagnetic field
can induce certain topological phase transition and the Chern number can be changed qualitatively
during this phase transition in the presence of light. Moreover, Uchoa and Seo~\cite{Uchoa1704.08780}
have pointed out that the superconductor quantum critical point can be actualized by
electric field and strain effects.

It is worth pointing out that the fermion-fermion interactions have been insufficiently
taken into account in previous studies. In this circumstance, the information of correlated
physical properties that are closely pertinent to these locally four-fermion interactions, in particular
the stability of ground states, the distance of Dirac nodal points in the semimetal states,
the transport properties and so on~\cite{Sachdev1999Book,Altland2002PR,
Lee2006RMP,Neto2009RMP,Fradkin2010ARCMP,Hasan2010RMP,Sarma2011RMP,Qi2011RMP,
Kotov2012RMP}, may be partially neglected or cannot be fully captured especially in the
low-energy regime. Incorporating more potential ingredients of physical degrees
of freedom in low-energy regime
would be of significant help to improve the description of their low-energy behaviors~\cite{Sachdev1999Book,
Altland2002PR,Lee2006RMP,Neto2009RMP,Fradkin2010ARCMP,Sarma2011RMP,
Kotov2012RMP}.  It is therefore imperative to examine what the role is played by the contribution from
fermion-fermion interactions to these low-energy phenomena of physical
properties in the 2D SD materials.

Within this work, we concentrate on four primary sorts of
short-range four-fermion interactions as designated in Eq.~(\ref{Eq_S_eff}). Additionally,
impurities are well-known to be present in the real systems and usually can play an essential
role in modern condensed matter physics~\cite{Ramakrishnan1985RMP,Nersesyan1995NPB,Mirlin2008RMP},
which can give rise to a plenty of prominent phenomena in low-energy regime~\cite{Ramakrishnan1985RMP,
Nersesyan1995NPB,Mirlin2008RMP,Efremov2011PRB,Efremov2013NJP,Korshunov2014PRB,Fiete2016PRB,
Nandkishore2013PRB,Potirniche2014PRB,Nandkishore2017PRB,Roy2016PRB-2,Roy1604.01390,Roy1610.08973,Roy2016SR}.
In Fermi systems, there conventionally own three sensible types of impurities~\cite{Nersesyan1995NPB,Stauber2005PRB}
that usually be named as random chemical potential, random mass, or random gauge potential~\cite{Nersesyan1995NPB,Stauber2005PRB,Wang2011PRB,Wang2013PRB} depending
upon their distinct couplings with fermions presented in Eq.~(\ref{Eq_S_f-d}). It has been proved that
these impurities can drive a multitude of interesting and unusual behaviors of physical
properties in these fermionic systems as widely shown in previous studies~\cite{Sachdev1999Book,Altland2002PR,Novoselov2005Nature,Aleiner2006PRL,Aleiner2006PRL-2,
Lee2006RMP,Neto2009RMP,Fradkin2010ARCMP,Hasan2010RMP,Sarma2011RMP,Wang2011PRB,Wang2013PRB,
Kotov2012RMP,Lee1702.02742}. In order to collect more physical information and fully understand
unconventionally physical properties in the low-energy regime, we suggeste to
turn on the four-fermion interactions and assume the presence of certain
amount of impurities simultaneously. At this stage, an intriguing question is therefore naturally raised
whether the basic results in noninteracting case with clean limit can be revised or even qualitatively
changed by the interplay between these four-fermion interactions and impurities.
Unambiguously answering this question would be of great profit for us to
further understand and uncover the unique properties of 2D SD materials and
even instructive to explore new Dirac-like materials.

In this paper, we are going to treat all the four types of short-range
four-fermion interactions and three kinds of impurities on the same footing via adopting the
momentum-shell renormalization group (RG) approach~\cite{Wilson1975RMP,Polchinski9210046,Shankar1994RMP}.
In this respect, the effects of four-fermion interactions, impurities, and
their interplays can be fully and unbiasedly incorporated into our consideration.
After taking into account all one-loop corrections from the competition between four-fermion
interaction and impurity, the coupled flow equations
of all associated interaction parameters for both the pure four-fermion interactions and presence of
impurity are derived after practicing the standard RG
analysis~\cite{Wilson1975RMP,Polchinski9210046,Shankar1994RMP,Wang2011PRB}.
To proceed, we employ these coupled flow equations that determine the evolutions of interaction
parameters replying on the energy scales to investigate whether and how the
low-energy behaviors of 2D SD systems can be affected or revised compared to
their clean and noninteracting counterparts, and additionally potential phenomena
would be trigged.

After performing both theoretical and numerical analysis, we find that several interesting
results have been extracted from these evolutions of all the correlated interaction
parameters. First, all of four-fermion parameters, to one-loop level,
are irrelevant in the RG language~\cite{Shankar1994RMP,Huh2008PRB,Wang2011PRB}.
This means the contribution from four-fermion
interactions becomes less and less significant and finally vanishes at the
lowest-energy limit~\cite{Shankar1994RMP}. As a result, these energy-dependent
interaction parameters at clean limit cannot qualitatively change the low-energy
behaviors of the 2D SD system~\cite{Shankar1994RMP,Huh2008PRB,Wang2011PRB}.
Consequently, both the trivial insulating and nontrivial semimetal states
are very stable against the four-fermion interactions. However, in the case of presence
of both four-fermion interaction and impurity, we find that the fates of these interaction
parameters can be modified qualitatively under certain initial conditions due to the strong
between fermion-fermion interactions and impurities. To be concrete,
these irrelevant interaction parameters can be transferred to irrelevant relevant
couplings after taking into account one-loop corrections
from interplay between fermion-fermion interactions and impurity~\cite{Wilson1975RMP,Polchinski9210046,
Shankar1994RMP,Wang2011PRB}, which lead to the divergence of interaction
parameters and instabilities at the critical energy scale. This conventionally suggests
that some phase transition~\cite{Fradkin2009PRL,Vafek2012PRB,Vafek2014PRB,
Wang2017QBCP,Altland2006Book,Vojta2003RPP,Metzner2000PRL,Metzner2000PRB,Wang2014PRB,Chubukov2010PRB,
Chubukov2012ARCMP,Chubukov2016PRX}, in our system certain insulator-semimetal phase transition expected,
would be generated under certain conditions although the states are qualitatively
stable against solely four-fermion interactions. In addition, we are informed that
the distance of two Dirac nodal points in the semimetal phase is sensitive
to the interplay and competition between the four-fermion interaction and impurity.
Specifically, we find that the four-fermion interactions and impurity respectively
decrease and increase the distance of Dirac nodal points.  One may expect that the
revisions of distance of Dirac nodal point would affect the interplay between the gapless
excitations in the low-energy regime. Moreover, several non-Fermi liquid
behaviors that deviate from the properties of conventional Fermi liquids theory~\cite{Altland2002PR},
for instance the quasiparticle residue $Z_f$ and the density of states (DOS) of the quasiparticle,
are obviously displayed at the lowest-energy limit caused by the intimate interplay and competition between
four-fermion interaction and random gauge potential or random mass.

The rest of this paper is organized as follows. In Sec.~\ref{Sec_model}, we bring out
the microscopic model and derive our effective quantum field theory. The forthcoming is
Sec.~\ref{Sec_RG_clean} that we provide the RG transformations for momenta, energies
and fields. All the one-loop corrections to the coupling parameters at clean limit
are computed in this section. In Sec.~\ref{Sec_RG_imp}, we subsequently move to derive
the coupled flow equations of all related parameters by means of the standard RG analysis
in the presence of both four-fermion interaction and impurity. In Sec.~\ref{Sec_int_delta},
we examine the stability of the ground states of 2D SD systems against the four-fermion
interactions at clean limit. Sec.~\ref{Sec_int-imp} and Sec.~\ref{Sec_NFL} are
accompanied by studying the low-energy behaviors affected by the interplay between
four-fermion interaction and impurity. Finally, we present a short summary in Sec.~\ref{Sec_summary}.

\section{Model and effective theory}\label{Sec_model}

\subsection{Noninteracting model}

The noninteracting Hamiltonian describing low-energy electronic bands of a SD
material can be generally proposed as~\cite{Banerjee2009PRL,Delplace2010PRB,Saha2016PRB}
\begin{eqnarray}
\mathcal{H}_0(\mathbf{k})=\mathbf{d}(\mathbf{k})\cdot\sigma,\label{Eq_H_0}
\end{eqnarray}
with $\mathbf{d}(\mathbf{k})=(\alpha k^2_x-\delta,vk_y,0)$ and $\mathbf{k}=(k_x,k_y,0)$.
Here, the parameters $\alpha$, $v$, and $\delta$ describe the inverse of quasiparticle
mass along $x$, the Dirac velocity along $y$, and the energy gap, respectively. In the rest of
this paper, we restrict the focus on the 2D systems. In this respect, we would easily
arrive at the the energy eigenvalues from this noninteracting
Hamiltonian (\ref{Eq_H_0})~\cite{Saha2016PRB}
\begin{eqnarray}
E^{\pm}(\mathbf{k})=\pm\sqrt{(\alpha k^2_x-\delta)^2+v^2k^2_y},\label{Eq_alpha}
\end{eqnarray}
with the notations $\pm$ respectively corresponding to the conduction and valence bands.
There are in all three potential ground states~\cite{Banerjee2009PRL,Delplace2010PRB,
Huang2015PRB,Saha2016PRB}: (i) $\delta=0$, the linear dispersion for $k_y$ direction
and parabolical for $k_x$ direction with a gapless spectrum; (ii) $\delta<0$, a trivial
insulating phase with a nonzero energy gap; and (iii) $\delta>0$, a 2D SD semimetal state
with two gapless Dirac nodal points at $(\pm\sqrt{\frac{\delta}{\alpha}},0)$.

We subsequently can address the noninteracting effective action
after performing several conventional transformations,
\begin{eqnarray}
S_0&=&\int\frac{d\omega}{(2\pi)}\int\frac{d^2\mathbf{k}}{(2\pi)^2}\Psi^\dagger(i\omega,\mathbf{k})
\Bigl[-i\omega+(\alpha k^2_x-\delta)\sigma_1\nonumber\\ \nonumber\\
&&+vk_y\sigma_2\Bigr]\Psi(i\omega,\mathbf{k}),\label{Eq_S_0}
\end{eqnarray}
with the spinor $\Psi(i\omega,\mathbf{k})$ characterizing the low-energy excitations from the Dirac nodes.
The $\sigma_j,j=1,2,3$ corresponds to the Pauli matrices. In order to study the low-energy behaviors,
we need to include the fermionic interactions besides this free term.

\begin{figure}
\centering
\includegraphics[width=3.0in]{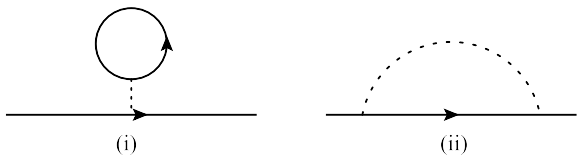}
\vspace{0.05cm}
\caption{One-loop corrections to the fermion propagator at clean limit (the dashed
line indicates the four-fermion interaction).}\label{Fig_fermion_propagator_correction}
\end{figure}

\subsection{Effective theory}

To proceed, we introduce the four-fermion interactions to construct our effective action.
After bringing out the four allowed kinds of short-range four-fermion interactions incorporated
into the non-interacting action (\ref{Eq_S_0})~\cite{Fradkin2009PRL,Vafek2012PRB,Vafek2014PRB,Wang2017QBCP},
we henceforth are left with the updated effective action
\begin{widetext}
\begin{eqnarray}
S_{\mathrm{eff}}&=&\int\frac{d\omega}{2\pi}\int\frac{d^2\mathbf{k}}{(2\pi)^2}\Psi^\dagger(i\omega,\mathbf{k})
[-i\omega+(\alpha k^2_x+\delta)\sigma_1+vk_y\sigma_2]\Psi(i\omega,\mathbf{k})+
\sum^3_{a=0}u_a\int\frac{d\omega_1d\omega_2d\omega_3}{(2\pi)^3}\int
\frac{d^2\mathbf{k}_1d^2\mathbf{k}_2d^2\mathbf{k}_3}{(2\pi)^{6}}\nonumber\\
&&\times\Psi^\dagger(\omega_1,\mathbf{k}_1)\sigma_a\Psi(\omega_2,\mathbf{k}_2)
\Psi^\dagger(\omega_3,\mathbf{k}_3)\sigma_a
\Psi(\omega_1+\omega_2-\omega_3,\mathbf{k}_1+\mathbf{k}_2-\mathbf{k}_3).\label{Eq_S_eff}
\end{eqnarray}
\end{widetext}
Here, the $\sigma_j,j=1,2,3$ again delineates the Pauli matrices and  $\sigma_0=I_{2\times2}$ the unit
matrix. The $u_a$, $a=0,1,2,3$ collects all sorts of given four-fermion interactions. This effective action
allows us to directly extract the free fermionic propagator,
\begin{eqnarray}
G_0(\omega,\mathbf{k})
&=&\frac{1}{-i\omega+\alpha k^2_x\sigma_1+vk_y\sigma_2}.\label{Eq_propagators}
\end{eqnarray}

To simplify our study and facilitate the evaluations, we can regard the $\delta$-term as
a quadratic interaction and its evolution upon lowering the energy scale will be attained in
the impending sections~\cite{Shankar1994RMP}. Based on these, we can expect the corresponding state
which the system locates at initially. In such circumstances, we are allowed to investigate how the
four-fermion interactions affect the low-energy behaviors of this 2D SD materials
via performing the momentum-shell RG analysis~\cite{Wilson1975RMP,Polchinski9210046,Shankar1994RMP}
of our effective theory~(\ref{Eq_S_eff}).

\section{One-loop RG analysis}\label{Sec_RG_clean}

\subsection{RG rescaling transformations}

According to the standard procedures of RG framework~\cite{Shankar1994RMP,Kim2008PRB,Huh2008PRB,
She2010PRB,She2015PRB,Roy2016PRB,Wang2011PRB,Wang2013PRB,Wang2013NJP,Vafek2012PRB,Wang2017QBCP,
Vafek2014PRB,Wang2014PRD,Wang2015PRB,Wang2017PRB}, we need to integrate out the fields
in the momentum shell $b\Lambda<k<\Lambda$ with $b<1$ to derive the evolutions of interaction
parameters, where $\Lambda$ represents the energy scale and the variable parameter $b$
can be written as $b=e^{-l}$ with a running energy scale $l>0$~\cite{Shankar1994RMP,
Kim2008PRB,Huh2008PRB,She2010PRB,She2015PRB,Wang2011PRB}.

Under the RG consideration, the parameters $\alpha$ and $v$ maybe flow upon lowering
the energy scale after collecting the one-loop interaction corrections. We will show later that the flows of
these two parameters would be revised and they enter into the coupled flow equations
of all interaction parameters after collecting the contribution from four-fermion interactions.
At this stage, all of interaction parameters are not independent but mutually and intimately
associated with others. Hence, the interacting couplings can also play a direct or indirect
role in the evolutions of the parameters $\alpha$ and $v$, which are essential to determine
the low-energy fate of 2D SD materials.

Before going to the one-loop RG calculations, we firstly dwell on the RG rescaling transformations.
In the spirt of momentum-shell RG theory~\cite{Shankar1994RMP,Huh2008PRB,She2010PRB,She2015PRB,Wang2011PRB},
we can choose the free action $S_0\sim-i\omega+\alpha k^2_x\sigma_1+vk_y\sigma_2$ as
the freely invariant fixed point that is invariant under the
RG transformation. As a result, we instantly address
the re-scaling transformations of momenta, energy and fermionic fields, namely,~\cite{Shankar1994RMP,Huh2008PRB,She2010PRB,She2015PRB,Wang2011PRB},
\begin{eqnarray}
k_{x}&=&k'_{x}e^{-\frac{1}{2}l},\label{Eq_RG_scales_k}\\
k_{y}&=&k'_{y}e^{-l},\label{Eq_RG_scales_k}\\
\omega&=&\omega'e^{-l},\label{Eq_RG_scales_omega}\\
\Psi(i\omega,\mathbf{k})&=&\Psi'(i\omega',\mathbf{k}')
e^{\frac{1}{2}\int_{0}^{l}dl\left(\frac{7}{2}-\eta_f\right)},\label{Eq_RG_scales_psi}
\end{eqnarray}
where the parameter $\eta_f$ delineates the anomalous fermion dimension~\cite{Shankar1994RMP,Huh2008PRB},
which would be generated by the four-fermion interaction
or impurities owning to their one-loop corrections.

\begin{figure}
\centering
\includegraphics[width=3.399in]{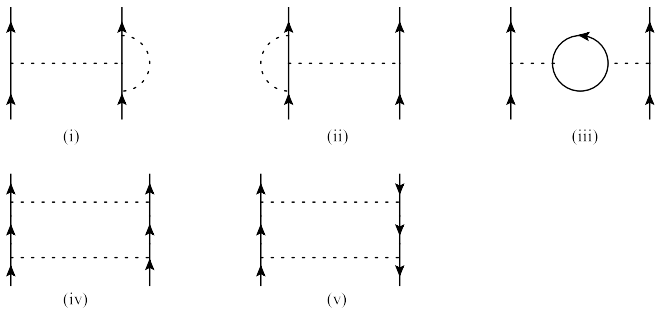}
\vspace{-0.3cm}
\caption{One-loop corrections to the four-fermion interacting couplings in clean limit (the dashed
line indicates the four-fermion interaction). All
subfigures (i)-(v) capture the contributions from purely four-fermion interactions distinguished by
different topological properties.}\label{Fig_fermion_interaction_correction}
\end{figure}

To proceed, we would like to present several clarifications on the RG re-scaling transformation
of energy $\omega$. It is indeed not unambiguous for us to choose its RG transformation
in that our microscopic model~(\ref{Eq_S_0}) owns a unconventional dispersion, namely,
linear in one direction and quadratical in the other, which is qualitatively distinct from two limit cases,
i.e. linear or quadratical in both $x$ and $y$ directions. Fortunately, it is worth highlighting that the
focus of present work is primarily on how the interaction parameters evolve upon lowering the energy
scale, in particular in the low-energy regime. This consequently suggests that the low-energy
degrees of freedom play a more significant role than their higher-energy counterparts.
In addition, learning from the noninteracting theory~(\ref{Eq_S_0}), the linear term $vk_y$ is
more dominant compared to the quadratical term $\alpha k^2_x$ while $k$ is adequate small
in the low-energy regime. These both indicate that the $z=1$ part (here $z$ depicts the dynamical
critical exponent) should be more important or take a leading responsibility for potentially
unique properties in the low-energy regime. Due to the exotic feature of dispersion
in the 2D SD material, one can expect the re-scaling transformation of $\omega$ (\ref{Eq_RG_scales_omega})
can properly capture the key ingredients of the low-energy physics. Therefore, we, in current work,
adopt the RG transformation of $\omega$, namely $\omega=\omega'e^{-l}$.

\subsection{Coupled flow equations}\label{Sec_RGeqs_clean}

In order to facilitate the performance of the standard momentum-shell RG analysis~\cite{Shankar1994RMP},
it is convenient to rescale the momenta and energy by $\Lambda_0$ that is linked to the lattice
constant, i.e. $k\rightarrow k/\Lambda_0$ and  $\omega\rightarrow\omega=\omega/\Lambda_0$,
and redefine the energy scale as $\Lambda=\Lambda_0/b$ with $b=e^{-l}$ and $l>0$ denoting
the changes of energy scales~\cite{Shankar1994RMP,Huh2008PRB,She2010PRB,Wang2011PRB}.
The free fermion propagator would receive one-loop corrections caused by the four-fermion
interaction as provided in Fig.~\ref{Fig_fermion_propagator_correction},
which can be explicitly expressed as
\begin{eqnarray}
\Sigma^{i}
&=&-\frac{u_1\mathcal{C}_1}{8\pi^2}\sigma_1l,\label{Eq_Sigma_i_clean}\\
\Sigma^{ii}&=&0,
\end{eqnarray}
with designating
\begin{eqnarray}
\mathcal{C}_1\equiv\int^{\frac{\pi}{2}}_{-\frac{\pi}{2}}d\theta\frac{2\alpha\cos\theta}
{\sqrt{\alpha^2\cos^2\theta+v^2\sin^2\theta}}.\label{Eq_C_1}
\end{eqnarray}
These one-loop corrections to the fermion propagator directly results in
\begin{eqnarray}
\eta_f=0,\label{Eq_eta_clean}
\end{eqnarray}
In this sense, we are straightforwardly informed that $v$ does not evolve via lowering
the energy scale~\cite{Shankar1994RMP,Roy2016PRB}, namely
\begin{eqnarray}
\frac{dv}{dl}&=&0,\label{Eq_RGeqs_v_clean}
\end{eqnarray}
and
\begin{eqnarray}
\frac{d\alpha}{dl}&=&0.\label{Eq_RGeqs_alpha_clean}
\end{eqnarray}
By virtue of the RG transformations (\ref{Eq_RG_scales_k})-(\ref{Eq_RG_scales_psi}) and
collecting the one-loop corrections (\ref{Eq_Sigma_i_clean}) and (\ref{Eq_RGeqs_v_clean}), we subsequently
arrive at the flow equation of parameter $\delta$ after fulfilling the standard momentum-shell RG analysis~\cite{Shankar1994RMP,Huh2008PRB,She2010PRB,Wang2011PRB},
\begin{eqnarray}
\frac{d\delta}{dl}&=&\left(1-\frac{u_1\mathcal{C}_1}{8\pi^2}\right)\delta.
\end{eqnarray}

\begin{figure}
\centering
\includegraphics[width=3.35in]{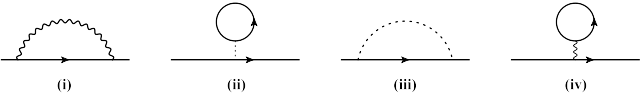}
\vspace{-0.105cm}
\caption{One-loop corrections to the fermion propagator (self-energy) in the presence of impurities
(The dashed and wave lines respectively indicate the four-fermion interaction and
the impurity).}\label{Fig_fermion_propagator_correction_2}
\end{figure}

Next, we move to address the evolutions of the fermion-fermion interaction couplings.
All one-loop corrections to the
four sorts of four-fermion interactions are depicted in Fig.~\ref{Fig_fermion_interaction_correction}
and the detailed results are presented in Appendix~\ref{Appendix_four-fermion-interaction}.
For instance, there are in all five one-loop diagrams contributing to the fermion interacting
coupling $u_0$ as shown in Fig.~\ref{Fig_fermion_interaction_correction},
whose contributions are displayed in Eqs.~(\ref{Eq_S_i_ii_u_0})-(\ref{Eq_S_v_u_0}) of Appendix~\ref{Appendix_four-fermion-interaction}.
Thereafter, summarizing all of these related one-loop contributions~\cite{Shankar1994RMP,Huh2008PRB,
She2010PRB,She2015PRB,Wang2011PRB,Wang2013NJP,Wang2014PRD,Wang2015PRB,Wang2017PRB} forthrightly gives
rise to
\begin{eqnarray}
&&\delta S_{u_0}\nonumber\\
&=&\int\frac{d\omega_1d\omega_2d\omega_3}{(2\pi)^3}\int^{b}
\frac{d^2\mathbf{k}_1d^2\mathbf{k}_2d^2\mathbf{k}_3}{(2\pi)^6}
\Psi^\dagger(\omega_1,\mathbf{k}_1)\sigma_0\Psi(\omega_2,\mathbf{k}_2)\nonumber\\
&&\times\Psi^\dagger(\omega_3,\mathbf{k}_3)\sigma_0\Psi(\omega_1+\omega_2
-\omega_3,\mathbf{k}_1+\mathbf{k}_2-\mathbf{k}_3)u_0\nonumber\\ \nonumber\\
&&\times \left[1-\frac{(u_2+u_3)(\mathcal{C}_2+\mathcal{C}_3-\mathcal{C}_0)
+u_1(3\mathcal{C}_2+\mathcal{C}_3-\mathcal{C}_0)}{8\pi^2}l\right]\nonumber\\ \nonumber\\
&\approx&\int\frac{d\omega_1d\omega_2d\omega_3}{(2\pi)^3}\int^{b}
\frac{d^2\mathbf{k}_1d^2\mathbf{k}_2d^2\mathbf{k}_3}{(2\pi)^6}
\Psi^\dagger(\omega_1,\mathbf{k}_1)\sigma_0\Psi(\omega_2,\mathbf{k}_2)\nonumber\\ \nonumber\\
&&\times\Psi^\dagger(\omega_3,\mathbf{k}_3)\sigma_0\Psi(\omega_1+\omega_2
-\omega_3,\mathbf{k}_1+\mathbf{k}_2-\mathbf{k}_3)\nonumber\\ \nonumber\\
&&\times u_0e^{\left[-\frac{(u_2+u_3)(\mathcal{C}_2+\mathcal{C}_3-\mathcal{C}_0)
+u_1(3\mathcal{C}_2+\mathcal{C}_3-\mathcal{C}_0)}{8\pi^2}l\right]}\nonumber\\ \nonumber\\
&=&\int\frac{d\omega_1d\omega_2d\omega_3}{(2\pi)^3}\int^1
\frac{d^2\mathbf{k}_1d^2\mathbf{k}_2d^2\mathbf{k}_3}{(2\pi)^6}\Psi^\dagger(\omega_1,\mathbf{k}_1)
\sigma_0\Psi(\omega_2,\mathbf{k}_2)\nonumber\\ \nonumber\\
&&\times\Psi^\dagger(\omega_3,\mathbf{k}_3)\sigma_0\Psi(\omega_1+\omega_2
-\omega_3,\mathbf{k}_1+\mathbf{k}_2-\mathbf{k}_3)\nonumber\\
&&\times u_0e^{\left[-1-\frac{(u_2+u_3)(\mathcal{C}_2+\mathcal{C}_3-\mathcal{C}_0)
+u_1(3\mathcal{C}_2+\mathcal{C}_3-\mathcal{C}_0)}{8\pi^2}\right]l},
\end{eqnarray}
which afterward results in the energy-dependent running equation of $u_0$ in the sprit
of momentum-shell RG theory~\cite{Shankar1994RMP,Huh2008PRB,She2010PRB,Wang2011PRB}
\begin{eqnarray}
\frac{du_0}{dl}
&=&\frac{u_0}{8\pi^2}\Bigl[(u_2+u_3)(\mathcal{C}_0-\mathcal{C}_2
-\mathcal{C}_3)\nonumber\\
&&+u_1(\mathcal{C}_0-3\mathcal{C}_2-\mathcal{C}_3)-4\pi^2\Bigr].
\end{eqnarray}

The RG flow equations of other coupling parameters can be deduced by paralleling above steps with employing
the corresponding one-loop corrections provided in Appendix~\ref{Appendix_four-fermion-interaction}.
After long but straightforwardly algebraic procedures, we eventually arrive at the coupled flow equations
of all related interaction parameters at clean limit~\cite{Shankar1994RMP,Huh2008PRB,She2010PRB,Wang2011PRB},
\begin{widetext}
\begin{eqnarray}
\frac{dv}{dl}&=&0,\label{Eq_RGeqs_alpha_clean}\\
\frac{d\alpha}{dl}&=&0,\label{Eq_RGeqs_alpha_clean}\\
\nonumber\\
\frac{d\delta}{dl}&=&\left(1-\frac{u_1\mathcal{C}_1}{8\pi^2}\right)\delta,\label{Eq_RGeqs_delta_clean}\\
\nonumber\\
\frac{du_0}{dl}&=&\frac{u_0(u_2+u_3)(\mathcal{C}_0-\mathcal{C}_2-\mathcal{C}_3)
+u_0u_1(\mathcal{C}_0-3\mathcal{C}_2-\mathcal{C}_3)-4\pi^2u_0}{8\pi^2},\label{Eq_RGeqs_u0_clean}\\
\nonumber\\
\frac{du_1}{dl}&=&\frac{u_1(u_0-u_2-u_3)(\mathcal{C}_0-\mathcal{C}_2)
+u_1(u_0-3u_2-u_3)\mathcal{C}_3-\mathcal{C}_2
(u^2_0+u^2_1+u^2_2+u^2_3)-4\pi^2u_1}{8\pi^2},\label{Eq_RGeqs_u1_clean}\\
\nonumber\\
\frac{du_2}{dl}&=&\frac{u_2(u_0-u_1-u_3)(\mathcal{C}_0+\mathcal{C}_2
-\mathcal{C}_3)-4\pi^2u_2}{8\pi^2},\label{Eq_RGeqs_u2_clean}\\
\nonumber\\
\frac{du_3}{dl}&=&\frac{u_3(u_0-u_1-u_2)(\mathcal{C}_0+\mathcal{C}_2
+\mathcal{C}_3)-4\pi^2u_3}{8\pi^2},\label{Eq_RGeqs_u3_clean}
\end{eqnarray}
\end{widetext}
where the three new coefficients $\mathcal{C}_i$, $i=0,2,3$ are representatively
defined by
\begin{eqnarray}
\mathcal{C}_0&=&\int^{\frac{\pi}{2}}_{-\frac{\pi}{2}}d\theta
\frac{2}{\sqrt{\alpha^2\cos^2\theta+v^2\sin^2\theta}},\label{Eq_C_0}\\
\mathcal{C}_2&=&\int^{\frac{\pi}{2}}_{-\frac{\pi}{2}}d\theta\frac{2\alpha^2\cos^2\theta}
{(\alpha^2\cos^2\theta+v^2\sin^2\theta)^{\frac{3}{2}}},\label{Eq_C_2}\\
\mathcal{C}_3&=&\int^{\frac{\pi}{2}}_{-\frac{\pi}{2}}d\theta\frac{2v^2\sin^2\theta}
{(\alpha^2\cos^2\theta+v^2\sin^2\theta)^{\frac{3}{2}}},\label{Eq_C_3}
\end{eqnarray}
with $\mathcal{C}_1$ being brought out in Eq.~(\ref{Eq_C_1}).

These coupled evolutions of
interaction parameters (\ref{Eq_RGeqs_alpha_clean})-(\ref{Eq_RGeqs_u3_clean}) indicate
that the four-fermion interacting couplings $u_a$, $a=0,1,2,3,4$ are pertinently
coupled and mutually influence each other via varying the energy scales. Their intimate
interplay may play a crucial role in determining the low-energy behaviors of physical
quantities, which will be carefully investigated and addressed in next section.

\section{One-loop RG analysis in the presence of impurity}\label{Sec_RG_imp}

In the real fermion systems, impurity is well
known to contribute significant effects to the low-energy behaviors of physical quantities~\cite{Ramakrishnan1985RMP,
Nersesyan1995NPB,Mirlin2008RMP} and consequently lead to a wealth of unconventional
phenomena in these Fermi systems~\cite{Ramakrishnan1985RMP,Nersesyan1995NPB,Mirlin2008RMP,
Efremov2011PRB,Efremov2013NJP,Korshunov2014PRB,Fiete2016PRB,
Nandkishore2013PRB,Potirniche2014PRB,Nandkishore2017PRB,Roy2016PRB-2,
Roy1604.01390,Roy1610.08973,Roy2016SR}. It is therefore of remarkable interest and
necessary to investigate how the impurity works before moving to examine how the
low-energy behaviors physical quantities are affected by the coupled flow equations (\ref{Eq_RGeqs_alpha_clean})-(\ref{Eq_RGeqs_u3_clean}) generated by the presence
of four-fermion interactions. Before going further, we would like to
suppose from now on that the semi-Dirac fermions are still extended in the presence
of weak impurity although there remains some debate on whether 2D Dirac/
semi-Dirac fermions extended or localized in the presence of impurity~\cite{Sarma2011RMP,Mirlin2008RMP}.
To proceed, we within this section endeavor to incorporate three important types of impurities
into our effective action (\ref{Eq_S_eff}), which are distinguished by their unique
couplings with fermions~\cite{Nersesyan1995NPB,Stauber2005PRB}, and named random
chemical potential, random mass, and random gauge potential respectively~\cite{Nersesyan1995NPB,Stauber2005PRB,Wang2011PRB,Wang2013PRB}.
Starting from the new effective theory incorporating the impurity, we extract
the revised version of coupled RG evolutions after collecting the interplay
between fermion and impurity.

\subsection{Fermion-impurity interaction}

The interplay between fermion and impurity can be conventionally described by the following expression~\cite{Ramakrishnan1985RMP,Nersesyan1995NPB,Stauber2005PRB,Mirlin2008RMP,Wang2011PRB}
\begin{eqnarray}
S_{\mathrm{fd}}=v_D\int d^2\mathbf{x}\Psi^\dagger(\mathbf{x})
\gamma\Psi(\mathbf{x})D(\mathbf{x}),\label{Eq_S_f-d}
\end{eqnarray}
where the Pauli matrix $\gamma$ represents different types of impurities and
$\gamma=\sigma_0$, $\gamma=\sigma_2$, and $\gamma=\sigma_{1,3}$ respectively
correspond to the random chemical potential, random mass and random gauge potential.
The coupling parameter $v_D$ denotes the impurity strength of a single impurity. The impurity field
$D(\mathbf{x})$ would be restricted by a quenched, Gauss-white potential under the conditions~\cite{Nersesyan1995NPB,Stauber2005PRB,Wang2011PRB,Altland2006Book,Coleman2015Book}
\begin{eqnarray}
\langle D(\mathbf{x})\rangle=0,\hspace{0.5cm}\langle D(\mathbf{x})D(\mathbf{x'})\rangle
=\Delta\delta^2(\mathbf{x}-\mathbf{x'}),\label{Eq_S_d-d}
\end{eqnarray}
here the $\Delta$ is assumed to measure the concentration of the impurity and can be taken as a constant
controlled by the experiment. After performing a Fourier transformation, we are left with the corresponding
action in momentum space~\cite{Nersesyan1995NPB,Stauber2005PRB,Wang2011PRB},
\begin{eqnarray}
S_{\mathrm{fd}}=v_D\int d^2\mathbf{k}d^2\mathbf{k'}d\omega\Psi^\dagger(\mathbf{k},\omega)
\gamma\Psi(\mathbf{k'},\omega)D(\mathbf{k-k'}).\label{Eq_S_f-d2}
\end{eqnarray}

\begin{figure}
\centering
\includegraphics[width=4.699in]{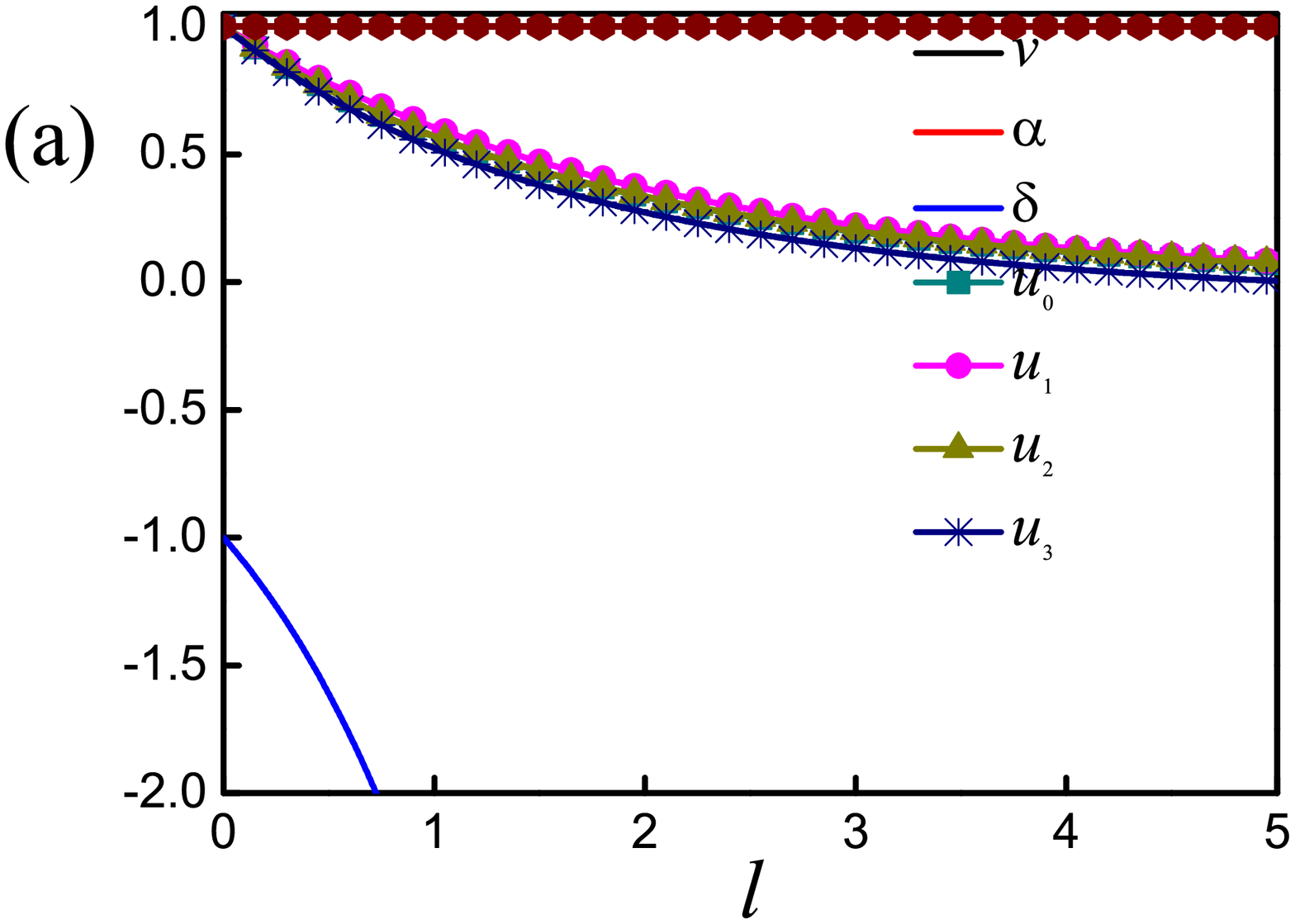}\vspace{-2.16cm}
\includegraphics[width=4.699in]{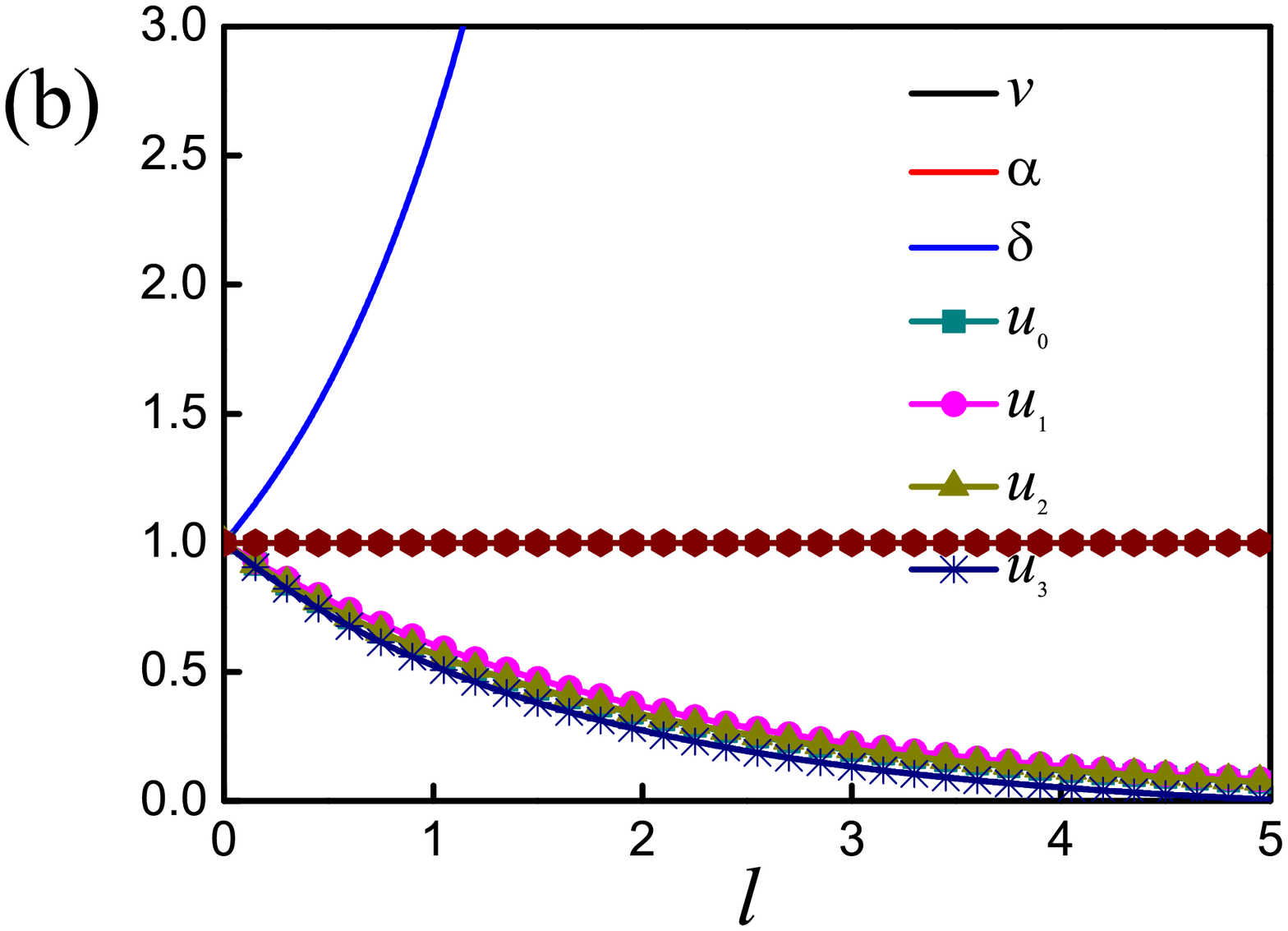}
\vspace{-2.23cm}
\caption{(Color online) Flows of interaction parameters due to the corrections
from the four-fermion interactions. The initial value of $\delta$ is taken to
be negative (a) or positive (b) and others equally.}\label{Fig_int_delta_M1}
\end{figure}

Based on these information, we eventually achieve the effective theory for the presence of impurity
by combing the effective action in Sec.~\ref{Sec_RG_clean} at clean limit and the fermion-impurity
interaction (\ref{Eq_S_f-d2}),
\begin{eqnarray}
S'_{\mathrm{eff}}&=&S_{\mathrm{eff}}+S_{\mathrm{fd}}.
\end{eqnarray}
Paralleling the one-loop analysis for fermion propagator in Sec.~\ref{Sec_RGeqs_clean} and computing
the one-loop fermion-impurity diagrams in Fig.~\ref{Fig_fermion_propagator_correction_2} for the presence
of random chemical potential, we find that the fermion propagator would gain a nontrivial revision
from the fermion-impurity interaction, namely
\begin{eqnarray}
\Sigma
&=&\frac{\Delta v^2_D\mathcal{D}_0}{8\pi^2}(i\omega)l.\label{Eq_Self_energy_0}
\end{eqnarray}
The other two sorts of fermion-impurity interactions both give rise to nonzero
corrections by practicing similar procedures,
\begin{eqnarray}
\Sigma
&=&-\frac{\Delta v^2_D\mathcal{D}_0}{8\pi^2}(i\omega)l.\label{Eq_Self_energy_123}
\end{eqnarray}

In such circumstance, we next can utilize these nontrivial corrections
to produce certain nonzero anomalous fermion dimensions, which are list as
\begin{eqnarray}
\eta^{\mathrm{chem}}_f&=&-\frac{\Delta v^2_D\mathcal{D}_0}{8\pi^2}=-\eta^{\mathrm{gaug}}_f
=-\eta^{\mathrm{mass}}_f.\label{Eq_eta}
\end{eqnarray}
where the indexes $\mathrm{chem},\mathrm{gaug},\mathrm{mass}$ denote the random chemical potential,
random gauge potential and random mass, representatively. These nonzero anomalous dimensions
will be intensely instructive to derive the coupled flow equations of interaction parameters
in next subsection, which are sharp contrast to
a trivial value at clean limit in Eq.~(\ref{Eq_eta_clean})~\cite{Nersesyan1995NPB,Huh2008PRB,Wang2011PRB}.
In addition, the impurity strength can participate in the coupled RG evolutions as a new member and
affect the revised evolutions of all correlated interaction parameters.

\subsection{Coupled flow equations in the presence of impurity}

After carrying out the analogous steps in section~\ref{Sec_RG_clean} with
employing the one-loop corrections in Appendix~\ref{Appendix_four-fermion-interaction_dis},
we consequently summarize the coupled flow equations of all related parameters
in the presence of random chemical potential,
\begin{widetext}
\begin{eqnarray}
\left.\begin{array}{ll}
\frac{dv}{dl}=\frac{\Delta v^2_D\mathcal{D}_0}{8\pi^2}v,\vspace{0.53cm}\\
\vspace{0.53cm}
\frac{d\alpha}{dl}=\frac{\Delta v^2_D\mathcal{D}_0}{8\pi^2}\alpha,\\
\vspace{0.53cm}
\frac{d\delta}{dl}=\left(1-\frac{u_1\mathcal{C}_1}{8\pi^2}+\frac{\Delta v^2_D\mathcal{D}_0}{8\pi^2}\right)\delta,\\
\vspace{0.53cm}
\frac{du_0}{dl}=\frac{2(\Delta v^2_D\mathcal{D}_0-2\pi^2)u_0
-u_0(u_2+u_3)(\mathcal{C}_2+\mathcal{C}_3-\mathcal{C}_0)
-u_0u_1(3\mathcal{C}_2+\mathcal{C}_3-\mathcal{C}_0)
-4\Delta v^2_D(\mathcal{D}_2+\mathcal{D}_3)}{8\pi^2},\\
\vspace{0.53cm}
\frac{du_1}{dl}=\frac{2(\Delta v^2_D\mathcal{D}_0-2\pi^2)u_1-u_1(u_2+u_3-u_0)(\mathcal{C}_0-\mathcal{C}_2)
-u_1(3u_2+u_3-u_0)\mathcal{C}_3-\mathcal{C}_2
(u^2_0+u^2_1+u^2_2+u^2_3)+2\Delta v^2_D(\mathcal{D}_2-\mathcal{D}_3)}{8\pi^2},\\
\vspace{0.53cm}
\frac{du_2}{dl}=\frac{2(\Delta v^2_D\mathcal{D}_0-2\pi^2)u_2-u_2(u_1+u_3-u_0)(\mathcal{C}_0+\mathcal{C}_2
-\mathcal{C}_3)+2\Delta v^2_D(\mathcal{D}_3-\mathcal{D}_2)}{8\pi^2},\\
\vspace{0.53cm}
\frac{du_3}{dl}=\frac{2(\Delta v^2_D\mathcal{D}_0-2\pi^2)u_3-u_3(u_1+u_2-u_0)(\mathcal{C}_0+\mathcal{C}_2
+\mathcal{C}_3)+6\Delta v^2_D(\mathcal{D}_2+\mathcal{D}_3)}{8\pi^2},\\
\vspace{0.53cm}
\frac{d v_D}{dl}=\frac{\Delta v^2_D\mathcal{D}_0+2[\Delta v^2_D-(u_0+u_1+u_2+u_3)]
(\mathcal{D}_2+\mathcal{D}_3)}{8\pi^2}v_D.
\end{array}\right. \label{Eq_RGeqs_int_dis_gamma0}
\end{eqnarray}
\end{widetext}
In order to present more compact evolutions of the interaction
parameters (\ref{Eq_RGeqs_int_dis_gamma0}),
we here have nominated four new coefficients that are
designated as
\begin{eqnarray}
\mathcal{D}_0&=&\int^{\frac{\pi}{2}}_{-\frac{\pi}{2}}d\theta\frac{2}{\alpha^2 \cos^2\theta+v^2\sin^2\theta},\label{Eq_D_0}\\
\mathcal{D}_1&=&\int^{\frac{\pi}{2}}_{-\frac{\pi}{2}}d\theta\frac{2\alpha\cos\theta}
{\alpha^2\cos^2\theta+v^2\sin^2\theta},\label{Eq_D_1}\\
\mathcal{D}_2&=&\int^{\frac{\pi}{2}}_{-\frac{\pi}{2}}d\theta\frac{2\alpha^2\cos^2\theta}
{(\alpha^2\cos^2\theta+v^2\sin^2\theta)^2},\label{Eq_D_2}\\
\mathcal{D}_3&=&\int^{\frac{\pi}{2}}_{-\frac{\pi}{2}}d\theta\frac{2v^2\sin^2\theta}
{(\alpha^2\cos^2\theta+v^2\sin^2\theta)^2}.\label{Eq_D_3}
\end{eqnarray}
The coupled flow equations for the presence of other two types of impurities
can derived analogously and are provided in the Appendix~\ref{Appendix_RGeqs_dis}.
In such situation, we can study the effects of these coupled flow equations that collect
contribution from both four-fermion interaction and impurity on the
low-energy behaviors of the semi-Dirac materials in the following section.

\section{Stability of the ground states against the fermionic interaction}\label{Sec_int_delta}

We have already presented the coupled RG flow equations of correlated coupling parameters
in Sec.~\ref{Sec_RG_clean} for the presence of fermion-fermion interactions. In the light
of these running Eqs.~(\ref{Eq_RGeqs_alpha_clean})-(\ref{Eq_RGeqs_u3_clean}), we are
informed that all these interaction parameters participate in the coupled evolutions
and mutually affect each other upon lowering energy scale. We subsequently concentrate
on the fate of parameter $\delta$ in low-energy regime under taking into account
one-loop interaction corrections.

Clearly, the parameter $\delta$ becomes energy-dependent and is coupled
with the evolutions of other parameters via reading the information provided in Eqs.~(\ref{Eq_RGeqs_alpha_clean})-(\ref{Eq_RGeqs_u3_clean}). To be specific,
the parameter $\delta$ flows via decreasing the energy scale and
it manifestly replies upon the coupling parameter $u_1$ and
additionally collects the contribution from the parameters $u_0$, $u_2$, and $u_3$
owning to their closely coupled RG equations. As told in Ref.~\cite{Saha2016PRB},
the parameter $\delta$ is extremely crucial to pin down
the ground state of the 2D SD system: (i) $\delta=0$, the spectrum is gapless with linear
dispersion along $k_y$; (ii) $\delta<0$, it is a gapped system with a trivial insulating phase; and
(iii) $\delta>0$, there exists two gapless Dirac nodal points at $(\pm\sqrt{\delta/\alpha},0)$
and this implies certain nontrivially topological state sets in. It is indeed that the system would
choose one of these states and be stable while we begin with the noninteracting
action~(\ref{Eq_H_0})~\cite{Saha2016PRB}.

However, we would like to stress that, the evolution of parameter $\delta$ with lowering the energy scale,
in apparently contrast to action (\ref{Eq_H_0}), is reconstructed profoundly while the four-fermion
interactions are switched on as presented in Eqs.~(\ref{Eq_RGeqs_alpha_clean})-(\ref{Eq_RGeqs_u3_clean}).
In addition,  all interaction parameters are not independent but
closely coupled with each other. In this respect, it is intensely
instructive to ask whether the low-energy properties of parameter $\delta$
can be qualitatively modified by incorporating into the one-loop corrections
due to the four-fermion interactions, namely whether these three ground states
are stable against these short-range four-fermion interactions?

In order to response to these questions, we have to numerically analyze the coupled RG
equations~(\ref{Eq_RGeqs_alpha_clean})-(\ref{Eq_RGeqs_u3_clean}). To this end, we
first take the initial value of parameter $\delta$ to be negative, i.e.,
$\delta(l=0)<0$ and rescale all the parameters by dividing their initial values, which are
supposed an equal starting constant for unbiased consideration.  After carrying out the
numerical calculations of coupled Eqs.~(\ref{Eq_RGeqs_alpha_clean})-(\ref{Eq_RGeqs_u3_clean})
with these initial conditions, we obtain interesting results as illustrated in Fig.~\ref{Fig_int_delta_M1}(a). This result directly suggests that the semimetal state with $\delta<0$ is insensitive to the four-fermion interaction and stable in the low-energy regime. Similarly, we representatively assume the parameter
$\delta$ to be positive initially and deliver the results after paralleling the
previous steps as designated in Fig.~\ref{Fig_int_delta_M1}(b). The qualitative results for the
case $\delta(l=0)=0$ are analogous and not shown here. According to the tendency of evolutions
upon decreasing the energy scale in Fig.~\ref{Fig_int_delta_M1}, we are unambiguously told that
the sign of parameter $\delta$ cannot be changed qualitatively by the fermion-fermion interactions.
This thereafter indicates both the trivial insulating state and nontrivial topological state are considerable
stable against the four-fermion interactions~\cite{Saha2016PRB}.

We stop here to present several comments on these
numerical results. At tree level, we easily obtain that all of the quartic interaction parameters
are irrelevant in the RG language after implementing the RG analysis of effective action, i.e.,  Eq.~(\ref{Eq_S_eff})~\cite{Shankar1994RMP,Huh2008PRB,Wang2011PRB}. In addition,
the numerical calculations of one-loop coupled flow equations countenance
this point as demonstrated in Fig.~\ref{Fig_int_delta_M1}. In the spirt of momentum-shell RG
theory~\cite{Shankar1994RMP}, this means that the fermion-fermion interaction parameters
are irrelevant even to one-loop level and as a result the contribution from four-fermion
interactions becomes less and less significant and finally vanishes at the
lowest-energy limit~\cite{Ramakrishnan1985RMP,Nersesyan1995NPB,Mirlin2008RMP}.
Consequently, these parameters cannot qualitatively destroy
the ground sates of the 2D SD material. 
Since the low-energy states of 2D SD system are insensitive
to these short-range four-fermion interaction, it is imperative to investigate
the effects of impurities, which are well known to be always
present in the real systems, and in particular interplay between
four-fermion interaction and impurity on these states.

\section{Low-energy behaviors affected by the interaction and impurity}\label{Sec_int-imp}

Impurities are well-known one of most significant facets in producing a multitude of prominent phenomena of
Fermi systems~\cite{Ramakrishnan1985RMP,Nersesyan1995NPB,Mirlin2008RMP}. Within this section,
we are going to study and answer whether our previously basic results with switching on the four-fermion
interactions at clean limit in the 2D SD system are robust under certain number of
impurities and their competition with fermion-fermion interaction.
In Fermi systems, there are three typical types of impurities, which are dubbed by
random chemical potential, random mass, and random gauge potential~\cite{Nersesyan1995NPB,Stauber2005PRB,
Wang2011PRB}. Without loss of generality, all these three sorts of impurities will be equally
investigated. After taking into account the contribution for the presence of both four-fermion
interaction and impurity, the coupled RG flow equations of interaction parameters are
modified from Eq.~(\ref{Eq_RGeqs_alpha_clean})-(\ref{Eq_RGeqs_u3_clean}) to
Eq.~\ref{Eq_RGeqs_int_dis_gamma0}, Eq.~(\ref{Eq_RGeqs_int_dis_gamma13}),
and Eq.~(\ref{Eq_RGeqs_int_dis_gamma2}).
Under such circumstances, we can expect the remarkably revised behavior
of parameter $\delta$ in the low-energy regime.

\subsection{Interaction-impurity induced phase transition}

To proceed, we endeavor to numerically calculate
Eq.~(\ref{Eq_RGeqs_int_dis_gamma0}), Eq.~(\ref{Eq_RGeqs_int_dis_gamma13}),
and Eq.~(\ref{Eq_RGeqs_int_dis_gamma2}) to check the energy-dependent properties
against the impurity. At the outset, we consider the presence of random chemical potential
with the coupled flow equations~(\ref{Eq_RGeqs_int_dis_gamma0}) and the other
two cases will be followed.

\begin{figure}
\centering
\includegraphics[width=4.699in]{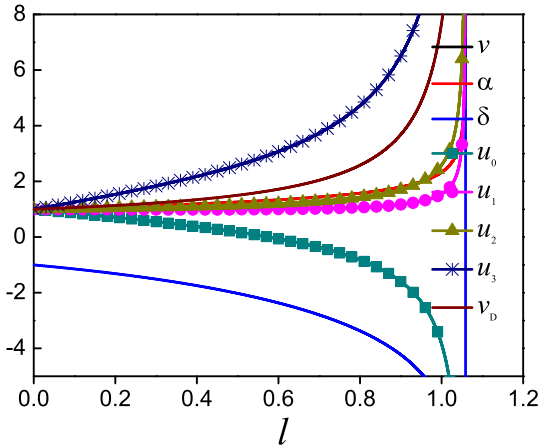}
\vspace{-2.16cm}
\caption{(Color online) Flows of parameters due to the corrections from the interplay between
four-fermion interactions and random chemical potential. The sign of parameter $\delta$ is changed
upon lowering the energy scale.}\label{Fig_int_imp_sign}
\end{figure}

In order to facilitate our analysis, we parallel the steps in Sec.~\ref{Sec_int_delta}
and initially let $\delta(l=0)=0$ and other parameters are assumed to own an equal starting value.
Before moving further, we elucidate the information on the strength of impurity.
For physical consideration, we introduce the impurity scattering rate to qualify the
initial value of impurity strength as~\cite{Coleman2015Book}
\begin{eqnarray}
\tau^{-1}\sim\frac{\Delta v^2_D}{\alpha},
\end{eqnarray}
where $\alpha$, $\Delta$ and $v_D$ are nominated in  Eqs.~(\ref{Eq_alpha}), (\ref{Eq_S_f-d})
and (\ref{Eq_S_d-d}), representatively. In the sprit of the perturbative theory,
we need to restrict the initial value of $\tau^{-1}(l=0)\equiv\tau^{-1}_0\leq\Lambda_0$.
After carrying out the similarly numerical performance, we find that the parameter
$\delta$ exhibits analogously compared to the pure-interaction case when the starting value
of impurity scattering rate is considerable small, namely $\tau^{-1}_0$ is much smaller than $\Lambda_0$.
In this case, the ground state of system is stable and the corresponding numerical results
are not shown here. On the contrary, while the initial value of impurity scattering rate is large,
for instance $\tau^{-1}_0>0.1\Lambda_0$, it is of particular interest that the sign of parameter
$\delta$ with a negative starting value can be changed upon lowering the energy scale as displayed in Fig.~\ref{Fig_int_imp_sign}.

This apparently signals that the system
undergoes a phase transition~\cite{Fradkin2009PRL,Vafek2012PRB,Vafek2014PRB,
Wang2017QBCP,Altland2006Book,Vojta2003RPP,Metzner2000PRL,Metzner2000PRB,Wang2014PRB,Chubukov2010PRB,
Chubukov2012ARCMP,Chubukov2016PRX}.  As shown in Fig.~\ref{Fig_int_imp_sign},
it is interesting to point out that the fermionic coupling parameters are divergent at
certain critical energy (dubbed by $l=l_c$). Guided by the spirt of momentum-shell RG~\cite{Wilson1975RMP,Polchinski9210046,Shankar1994RMP}
and phase transition theory~\cite{Sachdev1999Book,Altland2006Book,Vojta2003RPP}, we are informed,
compared to the clean-limit case with only switching the four-fermion interactions in Sec.~\ref{Sec_int_delta}, that the irrelevant fermion-fermion interaction parameters
can be transferred to irrelevant relevant interaction couplings in the low-energy
regime after collecting one-loop corrections due to the close interplay between
four-fermion interactions and impurity. Generally,
these irrelevant relevant parameters are responsible for the divergences of interaction parameters.
A multitude of previous studies~\cite{Fradkin2009PRL,Vafek2012PRB,Vafek2014PRB,
Wang2017QBCP,Altland2006Book,Vojta2003RPP,Metzner2000PRL,Metzner2000PRB,Wang2014PRB,Chubukov2010PRB,
Chubukov2012ARCMP,Chubukov2016PRX} reveal that these divergent coupling parameters
are the evident signals of instability and phase transitions. Consequently, we infer that the
intimate interplay between four-fermion interaction and impurity together triggers some phase transition
in the low-energy regime. At this stage, a good candidate for our 2D SD material occurs,
namely, a trivial insulator experiencing certain phase transition to become a nontrivial Dirac
semimetal owning to the strong interplay between random chemical potential and fermion-fermion interactions.

Subsequently, we move to the cases for presence of random gauge potential and random mass
via considering Eq.~(\ref{Eq_RGeqs_int_dis_gamma13}) and Eq.~(\ref{Eq_RGeqs_int_dis_gamma2})
in Appendix~\ref{Appendix_RGeqs_dis}. Paralleling the analysis for the random chemical potential
indicates these two sorts of impurities share the qualitative results with the random chemical
potential: while the impurity scattering rate is small, the ground states of 2D SD systems are
strongly stable against the interplay between fermion-fermion interaction and impurities;
however, the potential phase transition from a trivial insulator to a gapless Dirac nodal
system would be generated by the random gauge potential or random mass if the initial
value of scattering rate of impurity is adequate large, which is as the same order
as the value in Fig.~\ref{Fig_int_imp_sign}. The primary difference from the random
chemical potential is that critical energy scales $E_c$ at which certain phase transition
is generated are distributed into three distinct values. After performing both analogously
numerical and theoretical studies, we are left with the orders of critical energy scales as
\begin{eqnarray}
E^{\mathrm{chem}}_c<E^{\mathrm{mass}}_c<E^{\mathrm{gaug}}_c.
\end{eqnarray}
where $\mathrm{chem},\mathrm{mass},\mathrm{gaug}$ describes
the random chemical potential, random mass and random gauge potential, representatively.
As the numerical results are similar to the presence of random chemical potential,
we do not provide them here.

Based on these analysis, we conclude that the ground states of 2D SD materials
are qualitatively stable against all three types of impurities at a small impurity scattering
rate, but certain phase transition can be generated under specific
condition as exhibited in Fig.~\ref{Fig_int_imp_sign} while the initial value of
impurity scattering rate exceeds certain critical value to make the fermion-fermion
interaction parameters irrelevantly relevant and the interplay between four-fermion
interaction and impurity remarkably significant.

\begin{figure}
\centering
\includegraphics[width=4.699in]{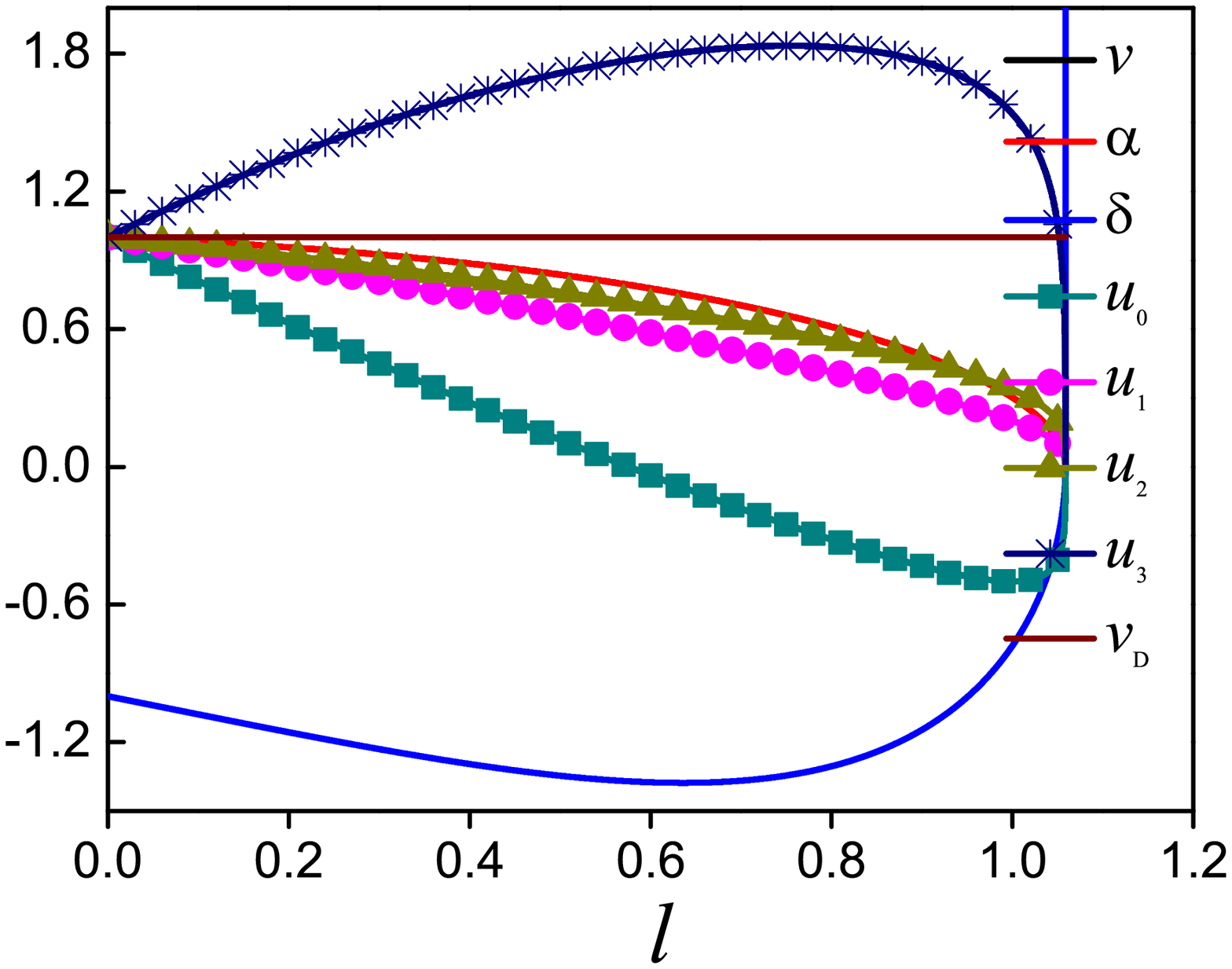}\vspace{-5.6cm}\\
\hspace{-1.3cm}\includegraphics[width=1.6in]{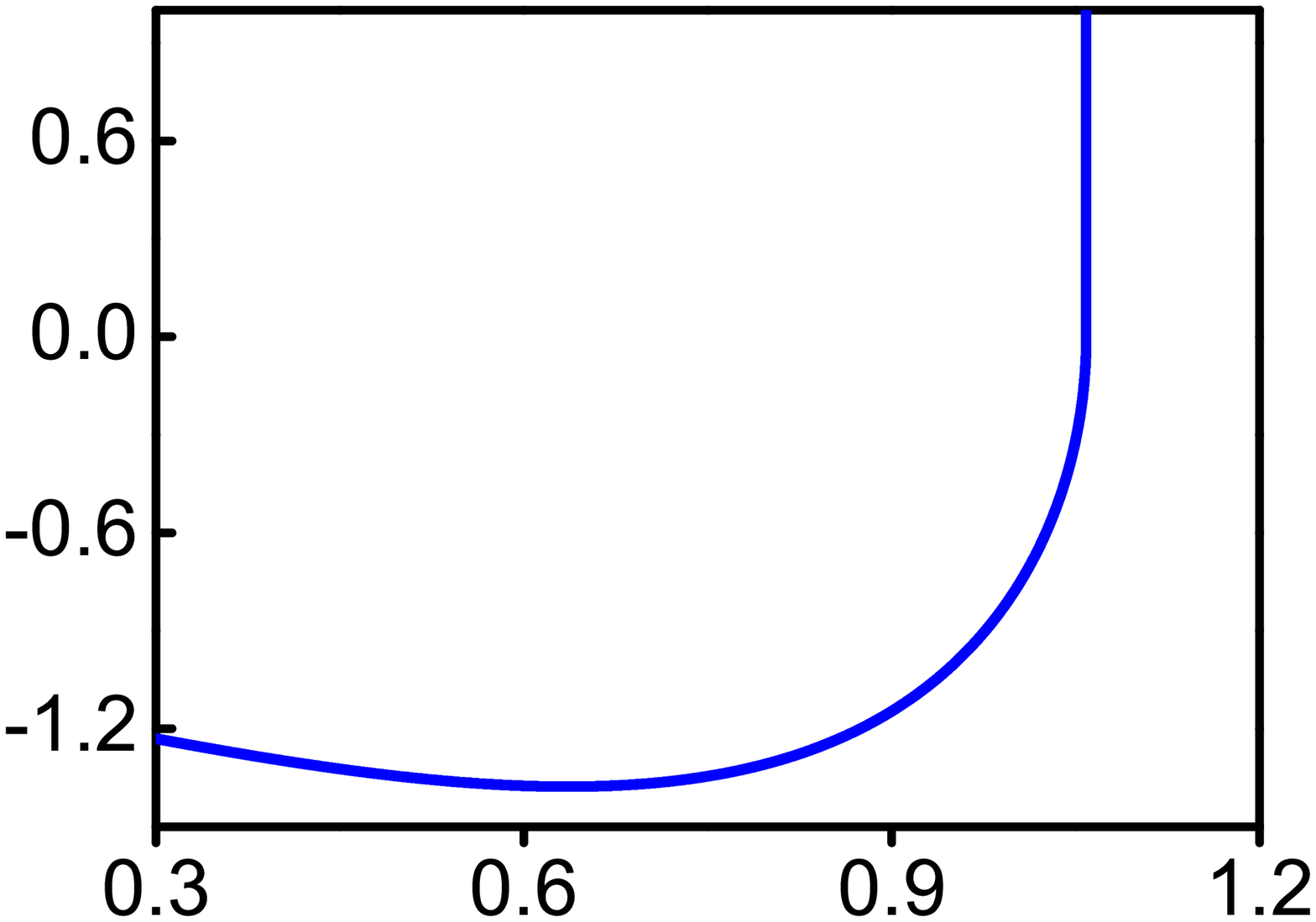}
\vspace{1.0cm}
\caption{(Color online) Relatively fixed points of flows of parameters due
to the corrections from the interplay between four-fermion interactions
and random chemical potential. The initial values are followed by the
Fig.~\ref{Fig_int_imp_sign} and the values of parameters are measured by
$v_D$. Inset: evolution of parameter $\delta$ in the sign-change regime.}\label{Fig_int_imp_sign_FP}
\end{figure}

\subsection{Relatively fixed point and phase transition}

Before closing the study of the impurity-induced phase transition,
we would like to provide our comments on this particular
phase transition, which is conventionally accompanied by the sign change of parameter $\delta$
shown in the last subsection.

To this end, we first need to seek the critical fixed point in the parameter space
that is lined to the critical energy scale represented by $l_c$ (or $E_c$) at
which the interaction couplings are divergent and instabilities take place~\cite{Shankar1994RMP,Vafek2012PRB,Vafek2014PRB,Metzner2000PRL,Metzner2000PRB,
Chubukov2010PRB,Chubukov2016PRX}. Learning from Fig.~\ref{Fig_int_imp_sign} informs us
that the interaction coupling parameters (or their absolute values) are increased
quickly and finally divergent at certain critical energy
scale. In order to make our analysis remain weak coupling, one can measure all of interaction couplings
with one of them (no sign change) and try to obtain the relatively fixed point of their
rations at this critical energy scale~\cite{Shankar1994RMP,Vafek2012PRB,Vafek2014PRB}.
Motivated by this idea and after performing analogously numerical analysis, we produce the relative
evolutions of interaction parameters and acquire our relatively
fixed point, namely, $[(v,\alpha,u_0,u_1,u_2,u_3,u_4,v_D)/v_D]^*
\approx(0,0,-5.0\times10^{-4},9.0\times10^{-5},2.4\times10^{-4},1.2\times10^{-3},1)$
by following the initial values of coupled flows in Fig.~\ref{Fig_int_imp_sign}.
In particular, we find that $[\delta/v_D]^*$ is substantially large and nearly divergent as
presented in Fig.~\ref{Fig_int_imp_sign_FP} in 2D SD system. Combing the evolutions of interaction
parameters in Fig.~\ref{Fig_int_imp_sign} and Fig.~\ref{Fig_int_imp_sign_FP},
it singles out that the divergence of parameter $\delta$ at the critical energy scale
should be intimately associated with the leading instability and simultaneously
accompanied by a phase transition. Given the perturbative restriction of our momentum-shell
RG framework~\cite{Shankar1994RMP}, we admit that our approach may gradually break
down and even be invalid at $l\geq l_c$ as the $\delta(l_c+\delta l)$ goes towards the strong coupling.
However, it is well proved that this analysis can conventionally provide constructive information
for leading instability and potential phase transition of the fermionic systems via approaching
the critical energy scale~\cite{Fradkin2009PRL, Vafek2012PRB,Vafek2014PRB,Wang2017QBCP,Altland2006Book,
Vojta2003RPP,Metzner2000PRL,Metzner2000PRB,Wang2014PRB,Chubukov2010PRB,Chubukov2012ARCMP,
Chubukov2016PRX}. Under all these consideration, we are encouraged to infer that some phase
transition must be trigged by the strong interplay between fermion-fermion interaction and
impurity in the vicinity of the critical energy scale denoted by $l_c$.

In order to judge which instability is dominant or which phase transition should be taken place,
we in principle need to calculate all of susceptibilities of potential instabilities allowed by
the symmetries of related system around its relatively fixed point and pick up the leading
one to be tied to the specific phase transition~\cite{Vafek2012PRB,Vafek2014PRB,Metzner2000PRL,
Metzner2000PRB,Chubukov2010PRB,Chubukov2016PRX}.

However, one usually, without loss of generality, can also
directly read the primary type of phase transition from the evolutions of interaction parameters~\cite{Chubukov2010PRB,Chubukov2012ARCMP}. Generally, the activated phase
transition is linked to the largest parameter at
the critical energy scale~\cite{Chubukov2010PRB,Chubukov2012ARCMP}.
As presented in Fig.~\ref{Fig_int_imp_sign} and Fig.~\ref{Fig_int_imp_sign_FP}, the parameter
$\delta$ features the largest value at the critical energy scale. Additionally,
the leading parameter $\delta$ is forced to go towards (positively) strong coupling
at $l=l_c$. Hence one can expect it would be positive at the low-energy regime.
Moreover, it clearly exhibits the sign change of the parameter $\delta$ upon
approaching the critical energy scale as displayed in the inset figure of
Fig.~\ref{Fig_int_imp_sign_FP}. In present study, we only try to put our
focus on how the four-fermion interaction and impurity qualitatively influence
the stabilities of the ground states of 2D SD systems. Therefore for the sake
of simplicity we suggest that the 2D SD system experiences a phase transition from
a trivial insulator to a nontrivial Dirac semimetal according to the spirt from
Refs.~\cite{Chubukov2010PRB, Chubukov2012ARCMP} caused by the interplay between
random chemical potential and four-fermion interaction.


\subsection{Sign change and phase transition}

To further clarify the implication of sign change,
we, within this subsection, provide further discussions on the relationship between
the sign change of parameter $\delta$ and phase transition. As studied in
above two subsections, the $\delta=0$ corresponds to an unstable fixed point that is
suggested to be linked to the phase transition. It is worth pointing out that the key ingredient
producing the sign change of $\delta$ is the divergent quantum fluctuation
at the phase transition point.

To be specific, we would like to emphasize, for physical consideration, the sign change
of $\delta$ is not solely taken place but closely and intimately associated with
the phase transition point at which the instability sets in accompanied concomitantly with the divergent
susceptibilities at the critical energy scale. By approaching the
quantum critical point (of the phase transition), the quantum fluctuations become stronger and
stronger, and finally divergent at that point. These unusual behaviors enter into
the coupled flow equations via the one-loop corrections and then greatly influence the
value of right-hand side of flow equations~(\ref{Eq_RGeqs_int_dis_gamma0}) in the
vicinity of phase transition point, such as the parameter $\delta$'s equation. As a result, the
sign change and divergence of parameter $\delta$ are triggered simultaneously under certain initial
conditions of four-fermion parameters and impurity strength.

For the mathematical consideration, we, for convenience, dub the right hand of flow equation
of $\delta$ as the slope of the $\delta$ equation nominated by $S_\delta$ and assume the unstable
fixed point takes place at $l_c$. If one starts from a $\delta_0<0$ with $S_{\delta_0}>0$, then
$\delta$ gradually approaches the fixed point via lowering the energy scale. On the contrary,
$\delta$ goes to the strong coupling once we begin from a $\delta_0>0$ still with $S_{\delta_0}>0$.
This explicitly implies the single-direction tendency of a parameter can be realized and closely tied
to the sign of $S_\delta$. To be concretely, if the sign of $S_\delta$ is the same while an infinitesimal
perturbation is tuned, $l_c\rightarrow l\pm\delta l$,
the sign change of unstable fixed point can be realized~\cite{Herbut2007Book}.
Guided by above discussion, we can straightforwardly numerically examine the $S_\delta$ away from
$l_c$ in current work (however the theoretical proof is hardly done for these complicatedly coupled
flow equations (\ref{Eq_RGeqs_int_dis_gamma0})). Before going further, we would like
to highlight that the "initial state" should be understood as a relative notation.
For instance, supposing the divergent behaviors set in at $l_c$,
we can define the "initial states" with $l^{\pm}_0\equiv l_c\pm\delta l$.
To be specific, the evolution can follow the steps: (i) initially,
$\delta$ starts from a negative value at $l^+_0$, and its absolute value
decreases in that the numerical calculation of coupled RG equations indicates $S_\delta(l^+_0)>0$;
(ii) while $l\rightarrow l_c$ it goes nearly zero, but the zero point is an unstable
fixed point impacted by the strong quantum fluctuation; (iii) then adding an infinitesimal
increase with $l^-_0$, the slope can change qualitatively by the divergent quantum
fluctuation still with $S_\delta(l^-_0>0)$ informed by the numerical evaluation,
which breaks the balance of the zero point fixed point and the parameter is  directly forced to
the strong coupling.

To reiterate, the intimate relationship between the evolutions of
coupled flow equations and strong quantum fluctuation yields to the singular behavior
of the $\delta$'s RG equation, and eventually results in the the sign change and divergence
of $\delta$ at phase transition point under certain initial conditions.

\begin{figure}
\centering
\includegraphics[width=4.699in]{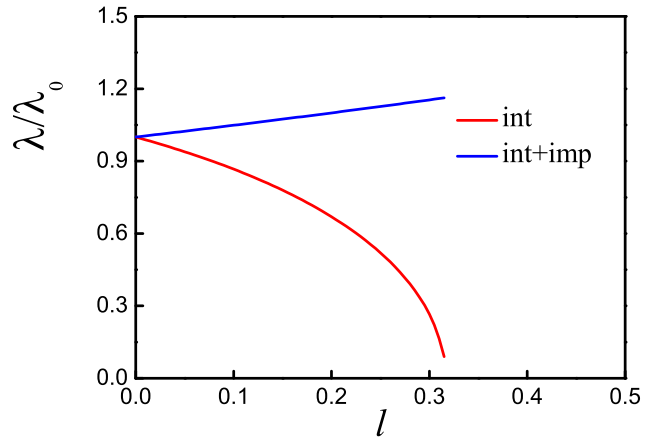}
\vspace{-2.3cm}
\caption{(Color online) Evolution of the distance ($\lambda/\lambda_0$)
between two nodal points in the $\delta>0$ states in the presence of pure
interaction ($\mathrm{int}$) and interaction+impruity ($\mathrm{int+imp}$)
before approaching the critical energy scale.}\label{Fig_int_imp_distance}
\end{figure}

\subsection{Distance of Dirac nodal points}

In addition to the phase transition investigated in former parts of Sec.~\ref{Sec_int-imp},
we here display another interesting behavior triggered by the interplay between four-fermion
interactions and impurity.

As addressed in Sec.~\ref{Sec_model}, the 2D SD material possesses two gapless Dirac nodal points
for $\delta>0$, namely at $(\pm\sqrt{\frac{\delta}{\alpha}},0)$.  Despite the gapless states with
$\delta>0$ being considerably stable against the short-range four-fermion interaction or impurity (assuming
the initial value of impurity scattering rate is not too large), the distance between these
two Dirac nodal points may be varied by the four-fermion interaction or impurity.
To proceed, their distance can be expressed as~\cite{Saha2016PRB}
\begin{eqnarray}
\lambda=2\sqrt{\frac{\delta}{\alpha}}.
\end{eqnarray}
This straightforwardly indicates that the distance $\lambda$ flows upon lowering
the energy scale as the correlated parameters both $\delta$ and $\alpha$ evolve and
are coupled complicatedly with other interaction parameters along the coupled RG flow equations~(\ref{Eq_RGeqs_alpha_clean})-(\ref{Eq_RGeqs_u3_clean})
or (\ref{Eq_RGeqs_int_dis_gamma0}), (\ref{Eq_RGeqs_int_dis_gamma13})
and (\ref{Eq_RGeqs_int_dis_gamma2}). We therefore are forced to
numerically analyze these coupled RG running equations to investigate the energy-dependent
properties of the distance of Dirac nodal points $\lambda$. Before moving further, we here would like to
stress that the following study is restricted to the energy scales which are larger than the
critical energy scale represented by $l_c$.

We subsequently take the starting value of $\delta$
positive to ensure the starting state harbors two Dirac nodal points and other parameters to
possess the similar values to Fig.~\ref{Fig_int_delta_M1}'s. The corresponding numerical results of
the energy-dependent distance for both cases are delineated in Fig.~\ref{Fig_int_imp_distance}.
We deduce that the four-fermion interaction can reduce the distance of Dirac nodal points after
capturing the basic information from Fig.~\ref{Fig_int_imp_distance}. On the contrary,
the presence all three sorts of impurities would increase the
distance $\lambda$ (as the difference among distinct types of impurities is negligible, we
only leave one line for impurity in Fig.~\ref{Fig_int_imp_distance}). As a consequence, we
conclude that the distance of Dirac nodal points in the semimetal states of 2D SD system
is sensitive to the four-fermion interaction or impurity, which representatively try to
decrease or increase the distance $\lambda$ as depicted in Fig.~\ref{Fig_int_imp_distance}.
One may expect that the increase of distance of Dirac nodal point would be somehow harmful to
the possibility of interplay between the gapless excitations in the low-energy regime.
Hence, this result is qualitatively consistent with the conclusion in Sec.~\ref{Sec_int_delta}
and Sec.~\ref{Sec_int-imp} that the ground state of 2D SD system is stable against
the four-fermion interaction and sensitive to the interplay between the
four-fermion interaction and impurity.

\section{Non-Fermi liquid behaviors}\label{Sec_NFL}

Based on our RG analysis in last several sections, the impurity indeed play a crucial role in
pinning down the physical quantities in the low-energy regime. We finally investigate
how the physical quantities behavior at the lowest-energy limit. In this section, we
focus on the quasiparticle residue $Z_f$ and DOS of the quasiparticle. As the random
chemical potential is a relevant parameter, our perturbative RG method
may break down at the lowest-energy limit and thus cannot clearly present an answer
to the fate of Landau quasiparticles~\cite{{Altland2002PR}}. Therefore, we here only
focus on the random gauge potential and random mass.

At the outset, we study the residue $Z_f$, which can be introduced via
\begin{eqnarray}
Z_f=\frac{1}{1-\frac{\partial \mathrm{Re} \Sigma^{\mathrm{R}}(\omega)}{\partial \omega}},
\end{eqnarray}
where $\mathrm{Re}\Sigma^{\mathrm{R}}$ represents the real part of retarded fermion self-energy.
It is worth pointing out that whether the notation of quasiparticle is well- or unwell- defined
(Fermi liquid or non-Fermi liquid) is closely associated with this renormalization factor $Z_{f}$.
On one side, a finite $Z_f$ corresponds to the Fermi liquid. On the other, the non-Fermi liquid (behavior) is usually accompanied by the residue $Z_f\rightarrow0$.
Subsequently, we can derive the evolution of $Z_f$ depending on the energy scale after performing
momentum-shell RG analysis and collecting the one-loop corrections to anomalous dimensions Eq.~(\ref{Eq_Self_energy_0})
and Eq.~(\ref{Eq_eta}), namely~\cite{Ludwig1994PRB,Xu2008PRB,WLZ2016NJP,Wang2017PRB2}
\begin{eqnarray}
\frac{d Z_f}{dl}=-\eta^i_f Z_f,
\end{eqnarray}
where $i=\mathrm{gaug},\mathrm{mass}$ as shown in Eq.~(\ref{Eq_eta}) representatively correspond
to the random gauge potential and random mass. By carrying the numerical analysis,
we find that $\lim_{l\rightarrow\infty}Z_f=0$ for both the random gauge potential and random mass,
this straightforwardly signals that the Dirac fermions are no longer well-defined
Landau quasiparticles.  As a result, non-Fermi liquid behaviors have been activated.

Additionally, we move to the DOS of the quasiparticle, whose flow $\rho$ under RG analysis can be
expressed as ~\cite{Ludwig1994PRB,WLZ2016NJP} for the 2D Dirac semimetal,
\begin{eqnarray}
\frac{d \ln\rho(\omega)}{d\ln(\omega)}=\frac{1+\eta^i_f}{1-\eta^i_f},\label{Eq_rho}
\end{eqnarray}
here again $i=\mathrm{gaug},\mathrm{mass}$ denotes distinct types of disorders.
This implies the DOS $\rho(\omega)\sim\omega$ for the clean and noninteracting 2D Dirac semimetal.
In order to examine the low-energy behaviors of the DOS in the presence of impurities,
we perform the numerical analysis after combining Eq.~(\ref{Eq_rho}) and coupled flow
equations for the presence of impurities, namely, Eq.~(\ref{Eq_RGeqs_int_dis_gamma13})
and Eq.~(\ref{Eq_RGeqs_int_dis_gamma2}) and subsequently are informed that the DOS at the lowest-energy
regime deviates from the linear behavior $\rho(\omega)\sim\omega$
due to the intimate interplay between four-fermion interactions
and random gauge potential or random mass. This is therefore also a signature
of non-Fermi liquid behavior.

\section{Summary}\label{Sec_summary}

In summary, we have carefully investigated the effects of four-fermion
interactions, impurities, and the interplay between fermionic interactions and
impurities on the ground states of 2D SD materials. In order to capture more
information of physical ingredients in the low-energy regime, all allowed types of
short-range four-fermion interactions by symmetries and three kinds of impurities,
namely random chemical potential, random mass, and random gauge potential~\cite{Nersesyan1995NPB,Stauber2005PRB,Wang2011PRB}, are
included and treated on the same footing by means of utilizing the momentum-shell RG theory~\cite{Wilson1975RMP,Polchinski9210046,Shankar1994RMP}.
The coupled flow equations for all related interaction parameters are derived after carrying
out the standard RG analysis~\cite{Wilson1975RMP,Polchinski9210046,Shankar1994RMP}
for both the pure interaction and presence of impurity, which are
employed to study the low-energy behaviors of 2D SD systems affected by the
interplay between the fermionic interaction and impurity. After performing both
analytic and numerical consideration of these coupled flow equations,
whether and how theses fermion-fermion interactions and impurity
impact the low-energy properties of the 2D SD materials are
fully studied and presented.

To be specific, we first consider the clean limit case with only switching on four sorts of
four-fermion interactions. After analyzing the evolutions of the
related interaction parameters upon lowering the energy scales, we clearly
display that all of four-fermion parameters
are irrelevant in the RG language~\cite{Shankar1994RMP,Huh2008PRB,Wang2011PRB}
to one-loop level. This indicates that the contribution from four-fermion
interactions becomes less and less significant and finally vanishes at the
lowest-energy limit~\cite{Shankar1994RMP}. Based on these theoretical and
numerical studies, we therefore conclude that both the trivial insulating
state and gapless semimetal state are considerably stable
against the short-range fermion-fermion interactions. However, while the impurity is
taken into account simultaneously, the close interplay between four-fermion interaction and impurity
can play a crucial role in determining the ground state of 2D SD system.
Concretely, these irrelevantly fermionic parameters can be transferred to
irrelevant relevant after collecting the corrections from interplay between interaction and impurity.
Additionally, if the initial value of impurity scattering exceeds certain critical
value, we find that the quartic interaction couplings are divergent and sign of parameter
$\delta$ can be changed concomitantly at a critical energy scale.
This conventionally suggests that some phase transition would
be triggered~\cite{Fradkin2009PRL,Vafek2012PRB,Vafek2014PRB,
Wang2017QBCP,Altland2006Book,Vojta2003RPP,Metzner2000PRL,Metzner2000PRB,Wang2014PRB,Chubukov2010PRB,
Chubukov2012ARCMP,Chubukov2016PRX}. For our 2D SD material, we expect
that the system experiences a phase transition from a trivial
insulator to a nontrivial Dirac, which is activated together by the interplay between four-fermion
interactions and impurities in the low-energy regime under specific conditions although the states are
qualitatively stable against all three types of impurities~\cite{Fradkin2009PRL,Vafek2012PRB,Vafek2014PRB,
Wang2017QBCP,Altland2006Book,Vojta2003RPP,Metzner2000PRL,Metzner2000PRB,Wang2014PRB,Chubukov2010PRB,
Chubukov2012ARCMP,Chubukov2016PRX}.

In addition, we find that the distance of two Dirac nodal points in
this 2D SD material is sensitive to the four-fermion interactions or impurities. On one side,
all three sorts of impurities are helpful to increase the distance of Dirac nodal points.
On the other, the four-fermion interaction is preferable to decreases it.
This interesting result is expected to be in line with the basic conclusion in Sec.~\ref{Sec_int_delta}
and Sec.~\ref{Sec_int-imp} that the ground state of 2D SD system is stable against
the four-fermion interaction and sensitive to the interplay between the
four-fermion interaction and impurity. Furthermore, several
deviations from Fermi liquid behaviors~\cite{Altland2002PR}, such as the quasiparticle residue $Z_f$
and the DOS, are displayed at the lowest-energy limit caused by the interplay between
the four-fermion interactions and random gauge potential or random mass. 
These exotic properties are manifest signals of non-Fermi liquid behaviors.

Before closing this section we would like to present brief comments on
the long-range fermion-fermion interaction. The long-range interactions may be possibly
formally rewritten as~\cite{Boettcher2016PRB,Hirata2016NComm}
\begin{eqnarray}
S_{\mathrm{int}}&=&
\sum^3_{a=0}u_a\int\frac{d\omega_1d\omega_2}{(2\pi)^2}\int
\frac{d^2\mathbf{k}_1d^2\mathbf{k}_2}{(2\pi)^{4}}
\Psi^\dagger(\omega_1,\mathbf{k}_1)\sigma_a\nonumber\\
&&\times\Psi(\omega_1,\mathbf{k}_1)
\Psi^\dagger(\omega_2,\mathbf{k}_2)\sigma_a
\Psi(\omega_2,\mathbf{k}_2),
\end{eqnarray}
which would give rise to totally different coupled evolutions of related parameters
in that one independent momentum and frequency would be vanished. Accordingly,
this interaction is expected to trigger more interesting behaviors.
However, one central problem is that the long-range
interactions are hardly present due to all kinds of screenings.
In particular, we take into account the effects of impurity scatterings, which
are unavoidably allowed in real systems and are very detrimental
for the long-range interactions. Under this circumstance,
it seems inappropriate to study the long-range interaction in this work, which
we leave for a future study hypothesizing the restricted conditions of presence of
long-range interaction being satisfied.

Finally, we expect that impeding experiments will be achieved to detect and verify whether there exists
the analogous phase transition from the gapped to gapless phases in the 2D SD materials, the variance
for distance of Dirac nodal points in the gapless phase, and also possibly non-Fermi
liquid behaviors at the low-energy regime. This may be profitable
for us to further understand and uncover the unique properties of 2D SD materials and even instructive to
motivate the new materials by virtue of many experimental methods~\cite{LCJS2009PRL,Beenakker2009PRL,
Rosenberg2010PRL,Hasan2011Science,Bahramy2012NC,Viyuela2012PRB,Bardyn2012PRL,Garate2003PRL,
Oka2009PRB,Lindner2011NP,Gedik2013Science,CBHR2016PRL}.


\section*{ACKNOWLEDGEMENTS}

J.W. is supported by the National Natural Science Foundation of China
under Grant No. 11504360.


\appendix

\section{One-loop corrections}

\subsection{Four-fermion couplings at clean limit}\label{Appendix_four-fermion-interaction}

According to Fig.~\ref{Fig_fermion_interaction_correction}(i) and (ii),
integrating out the fast modes of fermionic fields and carrying out the standard momentum-shell RG
framework~\cite{Shankar1994RMP,Huh2008PRB,She2010PRB,Wang2011PRB,Vafek2012PRB} by means of
utilizing the RG transformations of the momenta (\ref{Eq_RG_scales_k}),
energy (\ref{Eq_RG_scales_omega}), and fields (\ref{Eq_RG_scales_psi}) give rise to
the one-loop corrections at clean limit
\begin{widetext}
\begin{eqnarray}
\delta S^{i+ii}_{u_0}
&=&\left[\frac{-u_0(u_0+u_1+u_2+u_3)(\mathcal{C}_2+\mathcal{C}_3-\mathcal{C}_0)}{8\pi^2}l
\right]\int^{+\infty}_{-\infty}\frac{d\omega_1d\omega_2d\omega_3}{(2\pi)^3}\int^{b}
\frac{d^2\mathbf{k}_1d^2\mathbf{k}_2d^2\mathbf{k}_3}{(2\pi)^6}\nonumber\\ \nonumber\\
&&\times\Psi^\dagger(\omega_1,\mathbf{k}_1)\sigma_0
\Psi(\omega_2,\mathbf{k}_2)\Psi^\dagger(\omega_3,\mathbf{k}_3)\sigma_0\Psi(\omega_1+\omega_2
-\omega_3,\mathbf{k}_1+\mathbf{k}_2-\mathbf{k}_3)\label{Eq_S_i_ii_u_0},\\ \nonumber\\
\delta S^{iii}_{u_0}
&=&\left[\frac{u^2_0(\mathcal{C}_2+\mathcal{C}_3-\mathcal{C}_0)}{8\pi^2}l\right]
\int^{+\infty}_{-\infty}\frac{d\omega_1d\omega_2d\omega_3}{(2\pi)^3}\int^{b}
\frac{d^2\mathbf{k}_1d^2\mathbf{k}_2d^2\mathbf{k}_3}{(2\pi)^6}
\Psi^\dagger(\omega_1,\mathbf{k}_1)\sigma_0\Psi(\omega_2,\mathbf{k}_2)\nonumber\\ \nonumber\\
&&\times\Psi^\dagger(\omega_3,\mathbf{k}_3)
\sigma_0\Psi(\omega_1+\omega_2-\omega_3,\mathbf{k}_1+\mathbf{k}_2-\mathbf{k}_3)\label{Eq_S_iii_u_0},\\ \nonumber\\
\delta S^{iv}_{u_0}
&=&\left[\frac{\mathcal{C}_0(u^2_0+u^2_1+u^2_2+u^2_3)
-2\mathcal{C}_2u_0u_1-2\mathcal{C}_3u_0u_2}{16\pi^2}l\right]
\int^{+\infty}_{-\infty}\frac{d\omega_1d\omega_2d\omega_3}{(2\pi)^3}\int^{b}
\frac{d^2\mathbf{k}_1d^2\mathbf{k}_2d^2\mathbf{k}_3}{(2\pi)^6}\nonumber\\ \nonumber\\
&&\times\Psi^\dagger(\omega_1,\mathbf{k}_1)\sigma_0\Psi(\omega_2,\mathbf{k}_2)
\Psi^\dagger(\omega_3,\mathbf{k}_3)\sigma_0\Psi(\omega_1+\omega_2-\omega_3,\mathbf{k}_1
+\mathbf{k}_2-\mathbf{k}_3)\label{Eq_S_iv_u_0},\\ \nonumber\\
\delta S^{v}_{u_0}
&=&\left[\frac{-\mathcal{C}_0(u^2_0+u^2_1+u^2_2+u^2_3)
-2\mathcal{C}_2u_0u_1+2\mathcal{C}_3u_0u_2}{16\pi^2}l\right]
\int^{+\infty}_{-\infty}\frac{d\omega_1d\omega_2d\omega_3}{(2\pi)^3}\int^{b}
\frac{d^2\mathbf{k}_1d^2\mathbf{k}_2d^2\mathbf{k}_3}{(2\pi)^6}\nonumber\\ \nonumber\\
&&\times\Psi^\dagger(\omega_1,\mathbf{k}_1)\sigma_0\Psi(\omega_2,\mathbf{k}_2)
\Psi^\dagger(\omega_3,\mathbf{k}_3)\sigma_0\Psi(\omega_1+\omega_2-\omega_3,
\mathbf{k}_1+\mathbf{k}_2-\mathbf{k}_3),\label{Eq_S_v_u_0}
\end{eqnarray}
for one-loop corrections to $u_0$,
\begin{eqnarray}
\delta S^{i+ii}_{u_1}
&=&\left[\frac{u_1(u_0+u_1-u_2-u_3)(\mathcal{C}_0-\mathcal{C}_2+\mathcal{C}_3)}{8\pi^2}l\right]
\int^{+\infty}_{-\infty}
\frac{d\omega_1d\omega_2d\omega_3}{(2\pi)^3}\int^{b}\frac{d^2\mathbf{k}_1
d^2\mathbf{k}_2d^2\mathbf{k}_3}{(2\pi)^6}\nonumber\\ \nonumber\\
&&\times\Psi^\dagger(\omega_1,\mathbf{k}_1)\sigma_1\Psi(\omega_2,\mathbf{k}_2)
\Psi^\dagger(\omega_3,\mathbf{k}_3)\sigma_1\Psi(\omega_1+\omega_2
-\omega_3,\mathbf{k}_1+\mathbf{k}_2-\mathbf{k}_3),\label{Eq_S_i_ii_u_1}\\ \nonumber\\
\delta S^{iii}_{u_1}
&=&\left[\frac{u^2_1(\mathcal{C}_2-\mathcal{C}_3-\mathcal{C}_0)}{8\pi^2}l\right]
\int^{+\infty}_{-\infty}\frac{d\omega_1d\omega_2d\omega_3}{(2\pi)^3}\int^{b}
\frac{d^2\mathbf{k}_1d^2\mathbf{k}_2d^2\mathbf{k}_3}{(2\pi)^6}\nonumber\\ \nonumber\\
&&\times\Psi^\dagger(\omega_1,\mathbf{k}_1)\sigma_1\Psi(\omega_2,\mathbf{k}_2)
\Psi^\dagger(\omega_3,\mathbf{k}_3)\sigma_1\Psi(\omega_1+\omega_2-\omega_3,\mathbf{k}_1
+\mathbf{k}_2-\mathbf{k}_3),\label{Eq_S_iii_u_1}\\ \nonumber\\
\delta S^{iv}_{u_1}
&=&\left[\frac{2\mathcal{C}_0u_0u_1
-\mathcal{C}_2(u^2_0+u^2_1+u^2_2+u^2_3)-2\mathcal{C}_3u_1u_2}{16\pi^2}l\right]
\int^{+\infty}_{-\infty}\frac{d\omega_1d\omega_2d\omega_3}{(2\pi)^3}\int^{b}
\frac{d^2\mathbf{k}_1d^2\mathbf{k}_2d^2\mathbf{k}_3}{(2\pi)^6}\nonumber\\ \nonumber\\
&&\times\Psi^\dagger(\omega_1,\mathbf{k}_1)\sigma_1\Psi(\omega_2,\mathbf{k}_2)
\Psi^\dagger(\omega_3,\mathbf{k}_3)\sigma_1\Psi(\omega_1+\omega_2-\omega_3,\mathbf{k}_1
+\mathbf{k}_2-\mathbf{k}_3),\label{Eq_S_iv_u_1}\\ \nonumber\\
\delta S^{v}_{u_1}
&=&\left[\frac{-2\mathcal{C}_0u_0u_1
-\mathcal{C}_2(u^2_0+u^2_1+u^2_2+u^2_3)-2\mathcal{C}_3u_1u_2}{16\pi^2}l\right]
\int^{+\infty}_{-\infty}\frac{d\omega_1d\omega_2d\omega_3}{(2\pi)^3}\int^{b}
\frac{d^2\mathbf{k}_1d^2\mathbf{k}_2d^2\mathbf{k}_3}{(2\pi)^6}\nonumber\\ \nonumber\\
&&\times\Psi^\dagger(\omega_1,\mathbf{k}_1)\sigma_1\Psi(\omega_2,\mathbf{k}_2)
\Psi^\dagger(\omega_3,\mathbf{k}_3)\sigma_1\Psi(\omega_1+\omega_2-\omega_3,\mathbf{k}_1
+\mathbf{k}_2-\mathbf{k}_3),\label{Eq_S_v_u_1}
\end{eqnarray}
for one-loop corrections to $u_1$,
\begin{eqnarray}
\delta S^{i+ii}_{u_2}
&=&\left[\frac{u_2(u_0-u_1+u_2-u_3)(\mathcal{C}_0+\mathcal{C}_2
-\mathcal{C}_3)}{8\pi^2}l\right]\int^{+\infty}_{-\infty}
\frac{d\omega_1d\omega_2d\omega_3}{(2\pi)^3}\int^{b}\frac{d^2\mathbf{k}_1
d^2\mathbf{k}_2d^2\mathbf{k}_3}{(2\pi)^6}\nonumber\\ \nonumber\\
&&\times\Psi^\dagger(\omega_1,\mathbf{k}_1)\sigma_2
\Psi(\omega_2,\mathbf{k}_2)\Psi^\dagger(\omega_3,\mathbf{k}_3)\sigma_2\Psi(\omega_1+\omega_2
-\omega_3,\mathbf{k}_1+\mathbf{k}_2-\mathbf{k}_3),\label{Eq_S_i_ii_u_2}\\ \nonumber\\
\delta S^{iii}_{u_2}
&=&\left[\frac{u^2_2(\mathcal{C}_3-\mathcal{C}_2-\mathcal{C}_0)}{8\pi^2}l\right]
\int^{+\infty}_{-\infty}\frac{d\omega_1d\omega_2d\omega_3}{(2\pi)^3}\int^{b}
\frac{d^2\mathbf{k}_1d^2\mathbf{k}_2d^2\mathbf{k}_3}{(2\pi)^6}\nonumber\\ \nonumber\\
&&\times\Psi^\dagger(\omega_1,\mathbf{k}_1)\sigma_2\Psi(\omega_2,\mathbf{k}_2)
\Psi^\dagger(\omega_3,\mathbf{k}_3)
\sigma_2\Psi(\omega_1+\omega_2-\omega_3,\mathbf{k}_1+\mathbf{k}_2-\mathbf{k}_3),\label{Eq_S_iii_u_2}\\ \nonumber\\
\delta S^{iv}_{u_2}
&=&\left[\frac{2\mathcal{C}_0u_0u_2
-2\mathcal{C}_2u_1u_2-\mathcal{C}_3(u^2_0+u^2_1+u^2_2+u^2_3)}{16\pi^2}l\right]
\int^{+\infty}_{-\infty}\frac{d\omega_1d\omega_2d\omega_3}{(2\pi)^3}\int^{b}
\frac{d^2\mathbf{k}_1d^2\mathbf{k}_2d^2\mathbf{k}_3}{(2\pi)^6}\nonumber\\ \nonumber\\
&&\times\Psi^\dagger(\omega_1,\mathbf{k}_1)\sigma_2\Psi(\omega_2,\mathbf{k}_2)
\Psi^\dagger(\omega_3,\mathbf{k}_3)
\sigma_2\Psi(\omega_1+\omega_2-\omega_3,\mathbf{k}_1+\mathbf{k}_2-\mathbf{k}_3),\label{Eq_S_iv_u_2}\\ \nonumber\\
\delta S^{v}_{u_2}
&=&\left[\frac{-2\mathcal{C}_0u_0u_2
+2\mathcal{C}_2u_1u_2+\mathcal{C}_3(u^2_0+u^2_1+u^2_2+u^2_3)}{16\pi^2}l\right]
\int^{+\infty}_{-\infty}\frac{d\omega_1d\omega_2d\omega_3}{(2\pi)^3}\int^{b}
\frac{d^2\mathbf{k}_1d^2\mathbf{k}_2d^2\mathbf{k}_3}{(2\pi)^6}\nonumber\\ \nonumber\\
&&\times\Psi^\dagger(\omega_1,\mathbf{k}_1)\sigma_2\Psi(\omega_2,\mathbf{k}_2)
\Psi^\dagger(\omega_3,\mathbf{k}_3)
\sigma_2\Psi(\omega_1+\omega_2-\omega_3,\mathbf{k}_1+\mathbf{k}_2-\mathbf{k}_3),\label{Eq_S_v_u_2}
\end{eqnarray}
for one-loop corrections to $u_2$, and
\begin{eqnarray}
\delta S^{i+ii}_{u_3}
&=&\left[\frac{u_3(u_0-u_1-u_2+u_3)(\mathcal{C}_0+\mathcal{C}_2
+\mathcal{C}_3)}{8\pi^2}l\right]\int^{+\infty}_{-\infty}
\frac{d\omega_1d\omega_2d\omega_3}{(2\pi)^3}\int^{b}\frac{d^2\mathbf{k}_1
d^2\mathbf{k}_2d^2\mathbf{k}_3}{(2\pi)^6}\nonumber\\ \nonumber\\
&&\times\Psi^\dagger(\omega_1,\mathbf{k}_1)\sigma_3
\Psi(\omega_2,\mathbf{k}_2)\Psi^\dagger(\omega_3,\mathbf{k}_3)\sigma_3\Psi(\omega_1+\omega_2
-\omega_3,\mathbf{k}_1+\mathbf{k}_2-\mathbf{k}_3),\label{Eq_S_i_ii_u_3}\\ \nonumber\\
\delta S^{iii}_{u_3}
&=&\left[\frac{u^2_3(-\mathcal{C}_3-\mathcal{C}_2-\mathcal{C}_0)}{8\pi^2}l\right]
\int^{+\infty}_{-\infty}\frac{d\omega_1d\omega_2d\omega_3}{(2\pi)^3}\int^{b}
\frac{d^2\mathbf{k}_1d^2\mathbf{k}_2d^2\mathbf{k}_3}{(2\pi)^6}\nonumber\\ \nonumber\\
&&\times\Psi^\dagger(\omega_1,\mathbf{k}_1)\sigma_3\Psi(\omega_2,\mathbf{k}_2)
\Psi^\dagger(\omega_3,\mathbf{k}_3)
\sigma_3\Psi(\omega_1+\omega_2-\omega_3,\mathbf{k}_1+\mathbf{k}_2-\mathbf{k}_3),\label{Eq_S_iii_u_3}\\ \nonumber\\
\delta S^{iv}_{u_3}
&=&\left[\frac{(2\mathcal{C}_0u_0u_3
-2\mathcal{C}_2u_1u_3-2\mathcal{C}_3u_2u_3)}{16\pi^2}l\right]
\int^{+\infty}_{-\infty}\frac{d\omega_1d\omega_2d\omega_3}{(2\pi)^3}\int^{b}
\frac{d^2\mathbf{k}_1d^2\mathbf{k}_2d^2\mathbf{k}_3}{(2\pi)^6}\nonumber\\ \nonumber\\
&&\times\Psi^\dagger(\omega_1,\mathbf{k}_1)\sigma_3\Psi(\omega_2,\mathbf{k}_2)
\Psi^\dagger(\omega_3,\mathbf{k}_3)
\sigma_3\Psi(\omega_1+\omega_2-\omega_3,\mathbf{k}_1+\mathbf{k}_2-\mathbf{k}_3),\label{Eq_S_iv_u_3}\\ \nonumber\\
\delta S^{v}_{u_3}
&=&\left[\frac{(-2\mathcal{C}_0u_0u_3
+2\mathcal{C}_2u_1u_3-2\mathcal{C}_3u_2u_3)}{16\pi^2}l\right]
\int^{+\infty}_{-\infty}\frac{d\omega_1d\omega_2d\omega_3}{(2\pi)^3}\int^{b}
\frac{d^2\mathbf{k}_1d^2\mathbf{k}_2d^2\mathbf{k}_3}{(2\pi)^6}\nonumber\\ \nonumber\\
&&\times\Psi^\dagger(\omega_1,\mathbf{k}_1)\sigma_3\Psi(\omega_2,\mathbf{k}_2)
\Psi^\dagger(\omega_3,\mathbf{k}_3)
\sigma_3\Psi(\omega_1+\omega_2-\omega_3,\mathbf{k}_1+\mathbf{k}_2-\mathbf{k}_3),\label{Eq_S_v_u_3}
\end{eqnarray}
for one-loop corrections to $u_3$.

\begin{figure}
\centering
\includegraphics[width=5.5in]{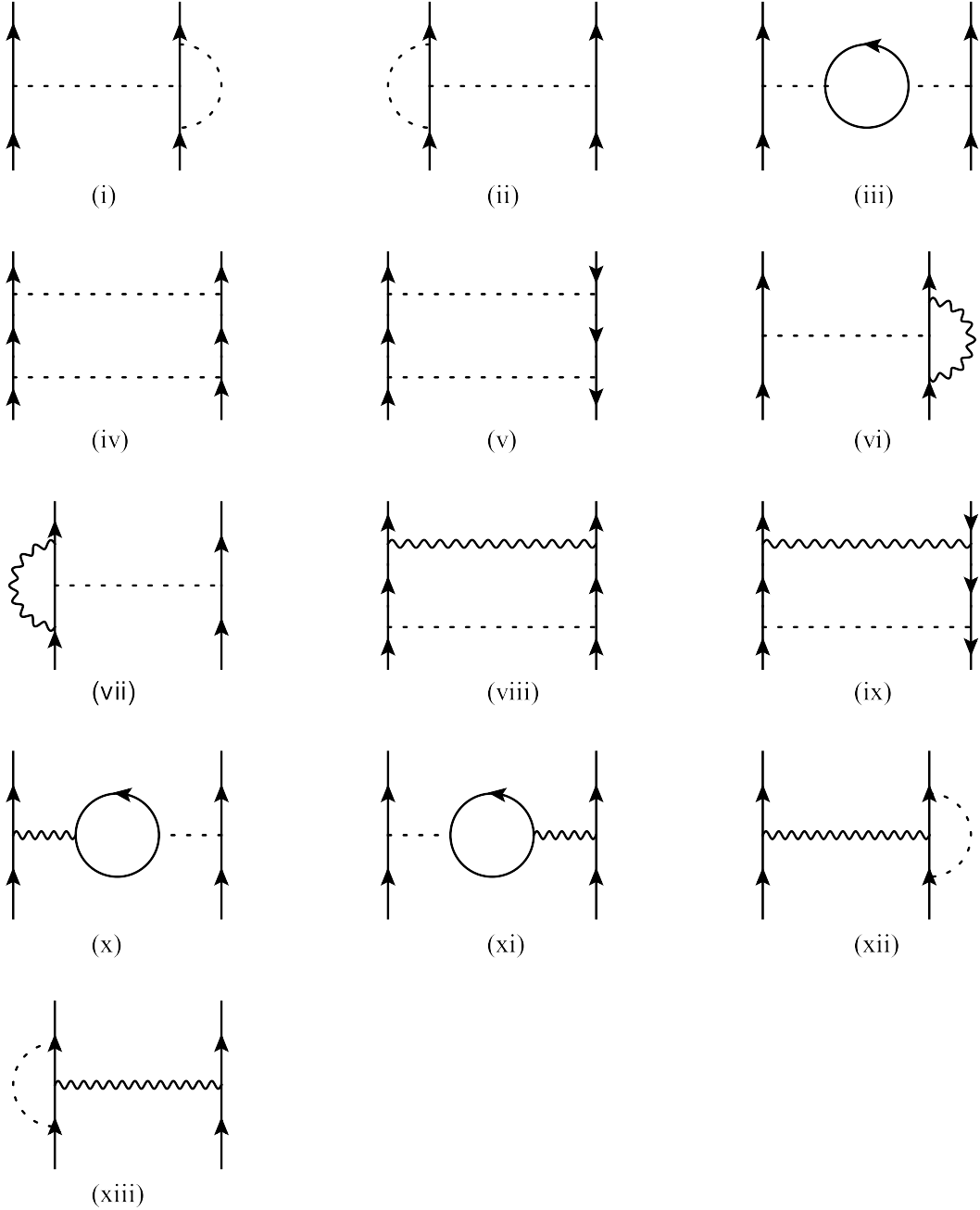}
\vspace{0.39cm}
\caption{One-loop corrections to the fermion interacting couplings in the presence of impurities.
The dashed line indicates the four-fermion interaction and the wave line represents the impurity. fermion interaction and the wave line represents the impurity. To be specific, the subfigures (i)-(v)
and (vi)-(xiii) respectively capture the contributions from purely four-fermion interaction and
mixtures of both four-fermion interaction and impurity.}\label{Fig_fermion_interaction_correction_2}
\end{figure}

\subsection{Four-fermion couplings in the presence of impurity}\label{Appendix_four-fermion-interaction_dis}

Due to the interplay between fermion-fermion interaction and impurity, we also need to take into
account the corrections from fermion-impurity interaction besides the fermion-fermion interaction
contributions provided in Appendix~\ref{Appendix_four-fermion-interaction}. In the presence of impurities,
all one-loop Feynman diagrams contributing to the four-fermion couplings are demonstrated in Fig.~\ref{Fig_fermion_interaction_correction_2}. As the subfigures (i)-(v) of Fig.~\ref{Fig_fermion_interaction_correction_2}
have already been evaluated in Appendix~\ref{Appendix_four-fermion-interaction}, we subsequently
will focus on the left subfigures of Fig.~\ref{Fig_fermion_interaction_correction_2}
one by one, i.e, subfigures (vi)-(xiii). For the presence of random chemical potential,
we can obtain the one-loop corrections
\begin{eqnarray}
\delta S^{vi+vii}_{u_0,\gamma=\sigma_0}
&=&\left[\frac{u_0\Delta v^2_D}{4\pi^2}\Bigl(\mathcal{D}_2+\mathcal{D}_3\Bigr)l\right]
\int\frac{d\omega_1d\omega_2d\omega_3}{(2\pi)^3}\int^{b}\frac{d^2\mathbf{k}_1d^2\mathbf{k}_2
d^2\mathbf{k}_3}{(2\pi)^6}\nonumber\\ \nonumber\\
&&\times\Psi^\dagger(\omega_1,\mathbf{k}_1)\sigma_0\Psi(\omega_2,\mathbf{k}_2)
\Psi^\dagger(\omega_3,\mathbf{k}_3)\sigma_0
\Psi(\omega_1+\omega_2-\omega_3,\mathbf{k}_1+\mathbf{k}_2-\mathbf{k}_3),\\ \nonumber\\
\delta S^{x+xi}_{u_0,\gamma=\sigma_0}
&=&\left[\frac{-u_0\Delta v^2_D}{\pi^2}\Bigl(\mathcal{D}_2+\mathcal{D}_3\Bigr)l\right]
\int\frac{d\omega_1d\omega_2d\omega_3}{(2\pi)^3}\int^{b}\frac{d^2\mathbf{k}_1d^2\mathbf{k}_2
d^2\mathbf{k}_3}{(2\pi)^6}\nonumber\\ \nonumber\\
&&\times\Psi^\dagger(\omega_1,\mathbf{k}_1)\sigma_0\Psi(\omega_2,\mathbf{k}_2)
\Psi^\dagger(\omega_3,\mathbf{k}_3)\sigma_0
\Psi(\omega_1+\omega_2-\omega_3,\mathbf{k}_1+\mathbf{k}_2-\mathbf{k}_3),\\ \nonumber\\
\delta S^{xii+xiii}_{u_0,\gamma=\sigma_0}
&=&\left[\frac{u_0\Delta v^2_D}{4\pi^2}\Bigl(\mathcal{D}_2+\mathcal{D}_3\Bigr)l\right]
\int\frac{d\omega_1d\omega_2d\omega_3}{(2\pi)^3}\int^{b}\frac{d^2\mathbf{k}_1d^2\mathbf{k}_2
d^2\mathbf{k}_3}{(2\pi)^6}\nonumber\\ \nonumber\\
&&\times\Psi^\dagger(\omega_1,\mathbf{k}_1)\sigma_0\Psi(\omega_2,\mathbf{k}_2)
\Psi^\dagger(\omega_3,\mathbf{k}_3)\sigma_0
\Psi(\omega_1+\omega_2-\omega_3,\mathbf{k}_1+\mathbf{k}_2-\mathbf{k}_3),\\ \nonumber\\
\delta S^{vi+vii}_{u_1,\gamma=\sigma_0}
&=&\left[\frac{u_1\Delta v^2_D(\mathcal{D}_2-\mathcal{D}_3)}{4\pi^2}l\right]
\int\frac{d\omega_1d\omega_2d\omega_3}{(2\pi)^3}\int^{b}\frac{d^2\mathbf{k}_1d^2\mathbf{k}_2
d^2\mathbf{k}_3}{(2\pi)^6}\nonumber\\ \nonumber\\
&&\times\Psi^\dagger(\omega_1,\mathbf{k}_1)\sigma_1\Psi(\omega_2,\mathbf{k}_2)
\Psi^\dagger(\omega_3,\mathbf{k}_3)\sigma_1
\Psi(\omega_1+\omega_2-\omega_3,\mathbf{k}_1+\mathbf{k}_2-\mathbf{k}_3),\\ \nonumber\\
\delta S^{vi+vii}_{u_2,\gamma=\sigma_0}
&=&\left[\frac{u_2\Delta v^2_D(\mathcal{D}_3-\mathcal{D}_2)}{4\pi^2}l\right]
\int\frac{d\omega_1d\omega_2d\omega_3}{(2\pi)^3}\int^{b}\frac{d^2\mathbf{k}_1d^2\mathbf{k}_2
d^2\mathbf{k}_3}{(2\pi)^6}\nonumber\\ \nonumber\\
&&\times\Psi^\dagger(\omega_1,\mathbf{k}_1)\sigma_2\Psi(\omega_2,\mathbf{k}_2)
\Psi^\dagger(\omega_3,\mathbf{k}_3)\sigma_2
\Psi(\omega_1+\omega_2-\omega_3,\mathbf{k}_1+\mathbf{k}_2-\mathbf{k}_3),\\ \nonumber\\
\delta S^{vi+vii}_{u_3,\gamma=\sigma_0}
&=&\left[\frac{-u_3\Delta v^2_D}{4\pi^2}\Bigl(\mathcal{D}_2+\mathcal{D}_3\Bigr)l\right]
\int\frac{d\omega_1d\omega_2d\omega_3}{(2\pi)^3}\int^{b}\frac{d^2\mathbf{k}_1d^2\mathbf{k}_2
d^2\mathbf{k}_3}{(2\pi)^6}\nonumber\\ \nonumber\\
&&\times\Psi^\dagger(\omega_1,\mathbf{k}_1)\sigma_3\Psi(\omega_2,\mathbf{k}_2)
\Psi^\dagger(\omega_3,\mathbf{k}_3)\sigma_3
\Psi(\omega_1+\omega_2-\omega_3,\mathbf{k}_1+\mathbf{k}_2-\mathbf{k}_3),\\ \nonumber\\
\delta S^{viii}_{u_3,\gamma=\sigma_0}
&=&\left[\frac{u_3\Delta v^2_D}{4\pi^2}\Bigl(\mathcal{D}_2\mathrm{Tr}(\sigma_1\sigma_a)\mathrm{Tr}(\sigma_a\sigma_1)
+\mathcal{D}_3\mathrm{Tr}(\sigma_2\sigma_a)\mathrm{Tr}(\sigma_a\sigma_2)\Bigr)l\right]
\int\frac{d\omega_1d\omega_2d\omega_3}{(2\pi)^3}\int^{b}\frac{d^2\mathbf{k}_1d^2\mathbf{k}_2
d^2\mathbf{k}_3}{(2\pi)^6}\nonumber\\ \nonumber\\
&&\times\Psi^\dagger(\omega_1,\mathbf{k}_1)\sigma_3\Psi(\omega_2,\mathbf{k}_2)
\Psi^\dagger(\omega_3,\mathbf{k}_3)\sigma_3
\Psi(\omega_1+\omega_2-\omega_3,\mathbf{k}_1+\mathbf{k}_2-\mathbf{k}_3).
\end{eqnarray}

\begin{figure}
\centering
\includegraphics[width=5.5in]{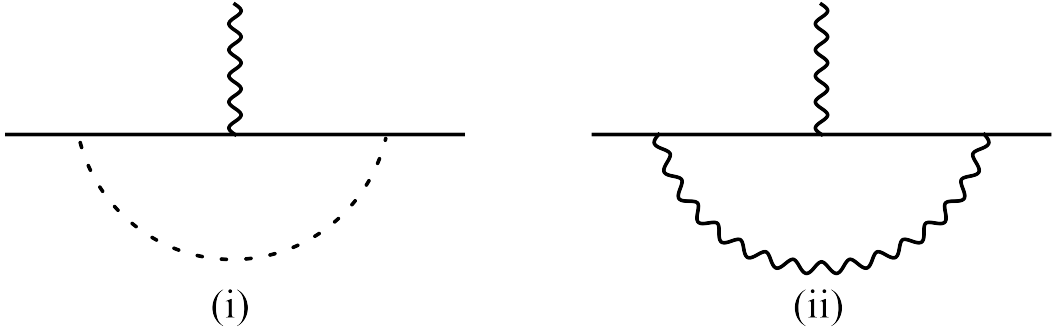}
\vspace{0.3cm}
\caption{One-loop corrections to the fermion-impurity interaction in the presence of impurities.
The dashed line indicates the four-fermion interaction and the wave line represents
the impurity.}\label{Fig_fermion_disorder_correction}
\end{figure}

We would like to emphasize that all of the subfigures in Fig.~\ref{Fig_fermion_interaction_correction_2}
that are not mentioned above have no any contributions. For the presence of random gauge potential and
random mass, one can derive the corrections similarly, which are not shown here.

\subsection{Fermion-impurity coupling}\label{Sec_one_loop_ferm_dis}

We finally study the one-loop corrections to the interplay between four-fermion interaction and
impurity, namely the fermion-impurity coupling, as illustrated in Fig.~\ref{Fig_fermion_disorder_correction}.
After carrying out analytical calculations in the spirt of momentum-shell RG approach~\cite{Shankar1994RMP,Huh2008PRB,She2010PRB,Wang2011PRB,Vafek2012PRB},
we list the results as follows for the presence of
all three types of impurities,
\begin{eqnarray}
\delta S_{\gamma=\sigma_0}
&=&\left\{\frac{v_D[\Delta v^2_D-(u_0+u_1+u_2+u_3)](\mathcal{D}_2+\mathcal{D}_3)}{4\pi^2}l\right\}
\int\frac{d\omega}{(2\pi)}\int^{b}\frac{d^2\mathbf{k}d^2\mathbf{k'}}{(2\pi)^4}\nonumber\\ \nonumber\\
&&\times\Psi^\dagger(\omega,\mathbf{k})\sigma_b\Psi(\omega,\mathbf{k'})D(\mathbf{k}-\mathbf{k'}),\\ \nonumber\\
\delta S_{\gamma=\sigma_1}
&=&\left\{\frac{v_D[\Delta v^2_D-(u_0+u_1-u_2-u_3)](\mathcal{D}_2
-\mathcal{D}_3)}{4\pi^2}l\right\}\int\frac{d\omega}{(2\pi)}
\int^{b}\frac{d^2\mathbf{k}d^2\mathbf{k'}}{(2\pi)^4}\nonumber\\ \nonumber\\
&&\times\Psi^\dagger(\omega,\mathbf{k})\sigma_b\Psi(\omega,\mathbf{k'})D(\mathbf{k}-\mathbf{k'}),\\ \nonumber\\
\delta S_{\gamma=\sigma_3}
&=&\left\{\frac{v_D[(u_0-u_1-u_2+u_3)-\Delta v^2_D](\mathcal{D}_2
+\mathcal{D}_3)}{4\pi^2}l\right\}\int\frac{d\omega}{(2\pi)}
\int^{b}\frac{d^2\mathbf{k}d^2\mathbf{k'}}{(2\pi)^4}\nonumber\\ \nonumber\\
&&\times\Psi^\dagger(\omega,\mathbf{k})\sigma_b\Psi(\omega,\mathbf{k'})D(\mathbf{k}-\mathbf{k'}),\\ \nonumber\\
\delta S_{\gamma=\sigma_2}
&=&\left\{\frac{v_D(u_0-u_1+u_2-u_3-\Delta v^2_D)(\mathcal{D}_2
-\mathcal{D}_3)}{4\pi^2}l\right\}
\int\frac{d\omega}{(2\pi)}\int^{b}\frac{d^2\mathbf{k}d^2\mathbf{k'}}{(2\pi)^4}\nonumber\\ \nonumber\\
&&\times\Psi^\dagger(\omega,\mathbf{k})\sigma_b\Psi(\omega,\mathbf{k'})D(\mathbf{k}-\mathbf{k'}),
\end{eqnarray}
where the Pauli matrix $\gamma=\sigma_0$, $\gamma=\sigma_2$, and $\gamma=\sigma_{1,3}$ respectively
correspond to the random chemical potential, random mass and random gauge potential.

\section{Coupled flow equations for the presence of impurities}\label{Appendix_RGeqs_dis}

By virtue of  paralleling the analysis of coupled RG equations in Sec.~\ref{Sec_RG_imp} for the random chemical potential,
the coupled flow equations for the presence of random gauge potential can be derived after performing
the standard procedures of momentum-shell RG framework~\cite{Shankar1994RMP,Huh2008PRB,She2010PRB,Wang2011PRB,Vafek2012PRB},
\begin{eqnarray}
\left.\begin{array}{ll}
\frac{dv}{dl}=-\frac{\Delta v^2_D\mathcal{D}_0}{8\pi^2}v, \vspace{0.539cm}\\
\vspace{0.539cm}
\frac{d\alpha}{dl}=\left(-1-\frac{\Delta v^2_D\mathcal{D}_0}{8\pi^2}\right)\alpha,\\
\vspace{0.539cm}
\frac{d\delta}{dl}=\left(1-\frac{u_1\mathcal{C}_1}{8\pi^2}-\frac{\Delta v^2_D\mathcal{D}_0}{8\pi^2}\right)\delta,\\
\vspace{0.539cm}
\frac{du_0}{dl}
=\frac{-2(\Delta v^2_D\mathcal{D}_0+4\pi^2)u_0-u_0(u_2+u_3)
(\mathcal{C}_2+\mathcal{C}_3-\mathcal{C}_0)
-u_0u_1(3\mathcal{C}_2+\mathcal{C}_3-\mathcal{C}_0)
+2\Delta v^2_D(2\mathcal{D}_2+\mathcal{D}_3)}{8\pi^2},\\
\vspace{0.539cm}
\frac{du_1}{dl}=\frac{-2(\Delta v^2_D\mathcal{D}_0+4\pi^2)u_1
-u_1(u_2+u_3-u_0)(\mathcal{C}_0-\mathcal{C}_2)
-u_1(3u_2+u_3-u_0)\mathcal{C}_3-\mathcal{C}_2
(u^2_0+u^2_1+u^2_2+u^2_3)+2\Delta v^2_D(2\mathcal{D}_3-\mathcal{D}_2)}{8\pi^2},\\
\vspace{0.539cm}
\frac{du_2}{dl}=\frac{-2(\Delta v^2_D\mathcal{D}_0+4\pi^2)u_2
-u_2(u_1+u_3-u_0)(\mathcal{C}_0+\mathcal{C}_2
-\mathcal{C}_3)+\Delta v^2_D(3\mathcal{D}_2-2\mathcal{D}_3)}{8\pi^2},\\
\vspace{0.539cm}
\frac{du_3}{dl}=\frac{-2(\Delta v^2_D\mathcal{D}_0+4\pi^2)u_3
-u_3(u_1+u_2-u_0)(\mathcal{C}_0+\mathcal{C}_2
+\mathcal{C}_3)+2\Delta v^2_D(3\mathcal{D}_2+2\mathcal{D}_3)}{8\pi^2},\\
\vspace{0.539cm}
\frac{d v_D}{dl}=\frac{2[(u_3-u_1)\mathcal{D}_2
+(u_0-u_2-\Delta v^2_D)\mathcal{D}_3]-\Delta v^2_D\mathcal{D}_0}{8\pi^2}v_D,
\end{array}\right.\label{Eq_RGeqs_int_dis_gamma13}
\end{eqnarray}
and finally carrying out analogous steps gives rise to the energy-dependent RG evolutions for presence of random mass,
\begin{eqnarray}
\left.\begin{array}{ll}
\frac{dv}{dl}=-\frac{\Delta v^2_D\mathcal{D}_0}{8\pi^2}v,\vspace{0.539cm}\\
\vspace{0.539cm}
\frac{d\alpha}{dl}=\left(-1-\frac{\Delta v^2_D\mathcal{D}_0}{8\pi^2}\right)\alpha,\\
\vspace{0.539cm}
\frac{d\delta}{dl}=\left(1-\frac{u_1\mathcal{C}_1}{8\pi^2}-\frac{\Delta v^2_D\mathcal{D}_0}{8\pi^2}\right)\delta,\\
\vspace{0.539cm}
\frac{du_0}{dl}
=\frac{-2(\Delta v^2_D\mathcal{D}_0+4\pi^2)u_0-u_0(u_2+u_3)
(\mathcal{C}_2+\mathcal{C}_3-\mathcal{C}_0)
-u_0u_1(3\mathcal{C}_2+\mathcal{C}_3-\mathcal{C}_0)
+2\Delta v^2_D(\mathcal{D}_2+\mathcal{D}_3)}{8\pi^2},\\
\vspace{0.539cm}
\frac{du_1}{dl}=\frac{-2(\Delta v^2_D\mathcal{D}_0+4\pi^2)u_1
-u_1(u_2+u_3-u_0)(\mathcal{C}_0-\mathcal{C}_2)
-u_1(3u_2+u_3-u_0)\mathcal{C}_3-\mathcal{C}_2
(u^2_0+u^2_1+u^2_2+u^2_3)+2\Delta v^2_D(\mathcal{D}_3-\mathcal{D}_2)}{8\pi^2},\\
\vspace{0.539cm}
\frac{du_2}{dl}=\frac{-2(\Delta v^2_D\mathcal{D}_0+4\pi^2)u_2
-u_2(u_1+u_3-u_0)(\mathcal{C}_0+\mathcal{C}_2
-\mathcal{C}_3)+6\Delta v^2_D(\mathcal{D}_2-\mathcal{D}_3)}{8\pi^2},\\
\vspace{0.539cm}
\frac{du_3}{dl}=\frac{-2(\Delta v^2_D\mathcal{D}_0+4\pi^2)u_3
-u_3(u_1+u_2-u_0)(\mathcal{C}_0+\mathcal{C}_2
+\mathcal{C}_3)+2\Delta v^2_D(\mathcal{D}_2+\mathcal{D}_3)}{8\pi^2},\\
\vspace{0.539cm}
\frac{d v_D}{dl}=\frac{2\left[(u_0-u_1+u_2-u_3)-\Delta v^2_D\right](\mathcal{D}_2
-\mathcal{D}_3)-\Delta v^2_D\mathcal{D}_0}{8\pi^2}v_D,
\end{array}\right.\label{Eq_RGeqs_int_dis_gamma2}
\end{eqnarray}
where the coefficients $\mathcal{C}_i$ and $\mathcal{D}_i$,
$i=0,1,2,3$ have been defined in the
maintext, namely Eqs.~(\ref{Eq_C_1}), (\ref{Eq_C_0})-(\ref{Eq_C_3}),
and Eqs.~(\ref{Eq_D_0})-(\ref{Eq_D_3}).
\end{widetext}


\end{document}